\documentclass[sigconf]{acmart}
\acmConference[EASE 2025]{The 29th International Conference on Evaluation and Assessment in Software Engineering}{17–20 June, 2025}{Istanbul, Türkiye}

\usepackage{pifont} % For the 3D star symbol
\usepackage{enumitem}
\usepackage{adjustbox}
\usepackage{subfigure}
\usepackage{graphicx}
\usepackage{multicol}
\usepackage{multirow}
\usepackage{xcolor,colortbl}
\usepackage[most]{tcolorbox}
\usepackage{textcomp}%
\usepackage{manyfoot}%
\usepackage{booktabs}%
\usepackage{adjustbox}
\usepackage{ragged2e}
\usepackage{array}
\usepackage{amsmath}
\usepackage{colortbl} % For cell color
\usepackage{tikz}     % For drawing circles

\newcommand{\threeDcircle}[1]{%
    \tikz\shade[ball color=#1] (0,0) circle (3pt);%
}

\definecolor{mycolor_box}{HTML}{F5FFFA} % Example Hex color F0FFFF E7FEFF FFFAF0 F0FFF0 E0FFFF
\definecolor{mycolor_title}{HTML}{FFEBCD} %FFEBCD FAE7B5

\AtBeginDocument{%
  }

\setcopyright{acmlicensed}
\copyrightyear{2025}
\acmYear{2025}
\acmDOI{XXXXXXX.XXXXXXX}
\acmISBN{978-1-4503-XXXX-X/18/06}

\begin{document}

%\title{Comparing Robustness Against Adversarial Attacks in Code Generation: LLM-Generated vs. Human-Written}

\title{Large Language Models as Robust Data Generators in Software Analytics: Are We There Yet?}

\author{Md. Abdul Awal}
\affiliation{
  \institution{University of Saskatchewan, Canada}
  \city{}
  \country{}
}
\email{mda439@usask.ca}
%\authornote{Corresponding author. Email: mda439@usask.ca (Md. Abdul Awal).}

\author{Mrigank Rochan}
\affiliation{
  \institution{University of Saskatchewan, Canada}
  \city{}
  \country{}
}
\email{mrochan@cs.usask.ca}

\author{Chanchal K. Roy}
\affiliation{
  \institution{University of Saskatchewan, Canada}
  \city{}
  \country{}
}
\email{chanchal.roy@usask.ca}

% %%
% %% By default, the full list of authors will be used in the page
% %% headers. Often, this list is too long, and will overlap
% %% other information printed in the page headers. This command allows
% %% the author to define a more concise list
% %% of authors' names for this purpose.

\begin{abstract}
%The rapid advancements in Large Language Models (LLMs) have led to the increasing use of LLM-generated data in various software analytics tasks, offering the potential to train high-performing models that complement human-created data. However, the comparative evaluation of human-written versus LLM-generated data for training high-performing models, especially under adversarial attacks in software analytics, remains largely unexplored.

Large Language Model (LLM)-generated data is increasingly used in software analytics, but it is unclear how this data compares to human-written data, particularly when models are exposed to adversarial scenarios. Adversarial attacks can compromise the reliability and security of software systems, so understanding how LLM-generated data performs under these conditions, compared to human-written data, which serves as the benchmark for model performance, can provide valuable insights into whether LLM-generated data offers similar robustness and effectiveness. To address this gap, we systematically evaluate and compare the quality of human-written and LLM-generated data for fine-tuning robust pre-trained models (PTMs) in the context of adversarial attacks. We evaluate the robustness of six widely used PTMs, fine-tuned on human-written and LLM-generated data, before and after adversarial attacks. This evaluation employs nine state-of-the-art (SOTA) adversarial attack techniques across three popular software analytics tasks: clone detection, code summarization, and sentiment analysis in code review discussions. Additionally, we analyze the quality of the generated adversarial examples using eleven similarity metrics. Our findings reveal that while PTMs fine-tuned on LLM-generated data perform competitively with those fine-tuned on human-written data, they exhibit less robustness against adversarial attacks in software analytics tasks. Our study underscores the need for further exploration into enhancing the quality of LLM-generated training data to develop models that are both high-performing and capable of withstanding adversarial attacks in software analytics.

\end{abstract}




\maketitle

%\vspace{-1.13em}

\section{Introduction}
\label{Intro}

Since data is the fuel for training or fine-tuning state-of-the-art (SOTA) deep learning (DL) models, ensuring access to high-quality data is crucial for developing robust, deep learning-based automated solutions in software analytics. Traditionally, human-created datasets have been regarded as the gold standard for training or fine-tuning DL models due to their richness, structure, diversity, and semantic depth \cite{ratner2016data}. However, acquiring large amounts of high-quality human-annotated data is expensive and time-consuming \cite{ratner2016data}. This limitation has led to the increasing use of LLM-generated data as a scalable and cost-effective alternative for various software analytics tasks \cite{shani_survey_2023}. For instance, Rahman et al. \cite{rahman2024words} augmented the existing ToxiCR \cite{sarker2023automated} dataset with LLM-generated data to improve the detection and translation of uncivil comments into civil ones in code review discussions. While LLM-generated data is increasingly used in software analytics, its quality and robustness compared to human-written data must be carefully assessed to ensure that models trained with this data are as effective and reliable, especially in adversarial scenarios.

Software analytics models trained on human-written data are vulnerable to adversarial attacks, leading to security risks, bug propagation, and technical debt in software maintenance and evolution \cite{yang2022natural, du2023extensive, guo2016exploring}. For instance, failing to detect code clones under adversarial attacks can result in technical debt and significant financial losses, estimated at $\$3.61$ per line of code \cite{guo2016exploring}. Given these challenges, it is crucial to investigate whether software analytics models trained on LLM-generated data exhibit similar vulnerabilities or potentially even more significant risks. To the best of our knowledge, this paper represents the first attempt in this direction. Such an investigation and comparison could provide valuable insights for improving the quality of LLM-generated data in software analytics, ultimately enhancing the security, robustness, and effectiveness of models in this domain.

While many studies have separately investigated the security, vulnerabilities, code smells, and robustness of data generated by LLMs \cite{cotroneo2023vulnerabilities, siddiq2024quality, wang2023adversarial, he2023large}, as far as we know, no research has yet directly examined how DL models trained on LLM-generated data perform compared to those trained on human-written data for the same software analytics tasks, particularly in terms of robustness under adversarial attacks. To fill this void, in this study, we systematically investigate the comparative quality of human-written and LLM-generated data for fine-tuning models in software analytics.

Our work is partly inspired by recent research in Natural Language Processing (NLP), where they compare human-written texts with those generated by LLMs for different tasks in NLP \cite{askari2023generating, guo2023close}. In our study, we utilize nine SOTA adversarial attack approaches (e.g., ALERT \cite{yang2022natural}, WIR-Random \cite{zeng2022extensive}, Metropolis-Hastings Modiﬁer (MHM) \cite{zhang2020generating}, BeamAttack \cite{du2023extensive}, TextFooler \cite{jin2020bert}, BAEAttack \cite{garg2020bae}, TextBugger \cite{li2018textbugger}, PWWSAttack \cite{ren2019generating} and BERT-Attack \cite{li2020bert}) to attack six widely-used PTMs (e.g., CodeBERT \cite{feng2020codebert}, CodeGPT \cite{lu2021codexglue}, PLBART \cite{ahmad2021unified}, BERT \cite{devlin2018bert}, RoBERTa \cite{liu2019roberta}, and DistilBERT \cite{sanh2019distilbert}). We perform an extensive study on three software analytics tasks, including clone detection, code summarization, and sentiment analysis in code review discussions. Our findings reveal that before adversarial attacks, while PTMs fine-tuned on LLM-generated data can achieve competitive performance in natural language-based tasks, PTMs fine-tuned on human-written data consistently outperform tasks requiring a nuanced understanding of program structure, such as clone detection. Furthermore, and more crucially, PTMs fine-tuned on human-written data exhibit greater resilience to adversarial attacks compared to those fine-tuned on LLM-generated data, generating adversarial examples that remain syntactically correct and semantically coherent, even after being modified by the attack. These findings emphasize the need for further research to enhance the quality of LLM-generated data, ensuring that models trained on it are both high-performing and resilient to adversarial attacks. To summarize, the main contributions of our study are listed below:

% Additionally, we advise exercising caution when using LLM-generated data for training or fine-tuning PTMs in real-world software analytics tasks, as PTMs fine-tuned on LLM-generated data tend to be more vulnerable to adversarial attacks.

\textbf{Originality.} To the best of our knowledge, this is the first comprehensive study to evaluate and compare the quality of human-written and LLM-generated data for robust model training in software analytics tasks, particularly under adversarial attacks.

% Our findings highlight the superiority of human-written data over LLM-generated data for training models that are more resilient to such attacks, emphasizing the need for further exploration to enhance the quality of LLM-generated data in this regard.

\textbf{Extensive study.} We systematically compare the performance of PTMs fine-tuned on human-written and LLM-generated datasets, considering six SOTA models, nine SOTA adversarial attack approaches, and three tasks: clone detection, code summarization, and sentiment analysis in code review discussions. Additionally, we extensively analyze the quality of adversarial examples generated from PTMs fine-tuned on human-written and LLM-generated data, utilizing eleven similarity measurement metrics.

\textbf{Dataset.} We preprocessed and filtered clone detection and sentiment analysis datasets to compare human-written and LLM-generated data. Additionally, we generated and validated 16K CodeSearchNet summaries using ASAP \cite{ahmed2024automatic} and SIDE \cite{mastropaolo2024evaluating}, establishing benchmarks for LLM-generated code summaries. Our preprocessed datasets provide benchmarks for evaluating human-written and LLM-generated data quality in future research.

\textbf{Open science} The code and the corresponding datasets used in our empirical study are publicly available \cite{Awal2025}.

\section{Background}
\label{back}

%This section briefly discusses some basic terminologies relevant to understanding this study.

\subsection{Pre-trained Models (PTMs)}
\label{PTMC}
Transformer-based \cite{vaswani2017attention} PTMs can learn general language representations from massive unlabeled corpora based on self-supervised objectives. Once trained, PTMs can be fine-tuned or adapted to new tasks using smaller datasets, thereby saving time and computational resources compared to training a model from scratch \cite{feng2020codebert, devlin2018bert}. The pre-training paradigm typically involves two main stages: \textit{pre-training} and \textit{fine-tuning}. In the pre-training stage, the model is trained on a large corpus of data using self-supervised learning techniques to learn general features or representations without requiring explicit labels \cite{feng2020codebert}. During the fine-tuning stage, the trained model adapts to a specific task with labeled examples by updating a smaller number of parameters using supervised learning techniques \cite{lu2021codexglue}. Therefore, fine-tuning not only streamlines the model training process but also dramatically reduces costs for a wide range of downstream tasks \cite{qiu2020pre}.

\subsection{Adversarial Attack on PTMs}

% \subsubsection{Software Analytics Tasks}
% This study evaluates the quality of LLM-generated synthetic data for fine-tuning PTMs, comparing it to human-written data in software analytics tasks. To facilitate this evaluation, we carefully select software analytics tasks where human-written and LLM-generated data are available to solve the same task. Additionally, to ensure the generalizability of our study findings, we consider diverse tasks in software analytics. This careful consideration of diverse tasks makes our study thorough and well-rounded. Consequently, we select three types of software analytics tasks: clone detection, code summarization, and sentiment analysis in code review discussions.

% \textbf{Code Summarization}. It is a task aimed at generating descriptions that explain the functionality of a given code snippet \cite{zhang2020retrieval}.

% \textbf{Clone Detection}. It is a task focused on determining whether two source code snippets are identical or similar \cite{roy2007survey}.

% \textbf{Sentiment Analysis}. It determines whether the comments in code review discussions are civic or non-civic \cite{sarker2023automated, rahman2024words}.

\subsubsection{Attack Definitions} Szegedy et al. \cite{szegedy2013intriguing} first introduced the notion of adversarial attack in computer vision. They showed that carefully crafted pixel-level perturbations imperceptible to the human eye can fool SOTA image classifiers. In discrete domains such as a clone detection task, we expect a PTM classifier $f: X \rightarrow Y$ to predict the test instance correctly (e.g., $y_{truth} \in Y$) for a code snippet $x \in X$. An adversarial attack aims to add slight perturbations to $x$ to generate an adversarial example $x'$ that can mislead $f$ following three requirements \cite{zhang2020generating, zeng2022extensive}: (1) Adversarial examples should mislead the PTMs as much as possible: $f(x') \neq f(x) = y_{truth}$. (2) Adversarial perturbations must maintain syntactic correctness, adhering to the syntax rules of the programming language. For example, in Python, an identifier name cannot start with a number after perturbation and must only contain letters, numbers, and underscores. (3) $x'$ must be semantically equivalent to $x$ after perturbations, meaning it has the same functionality and produces the same result for the same input. We follow the attack settings of existing studies \cite{jin2020bert, li2018textbugger, zeng2022extensive, du2023extensive} to generate adversarial examples that adhere to the constraints mentioned above. When subtle modifications following the above constraints are applied to an original input, resulting in a transformed version resembling the original, we refer to the changed version as an adversarial example. If the modified sample alters the prediction made by the PTM, it is categorized as an adversarial attack.

\subsubsection{Adversarial Robustness} It refers to the ability of a machine learning model to maintain its performance and make accurate predictions even in the presence of adversarial examples \cite{szegedy2013intriguing}. For instance, if a PTM model can perform well under adversarial attacks and maintain accuracy on such examples, we consider it robust against such attacks.

\section{Study Design}
\label{method}

\subsection{Datasets and Subjects}
\label{datasets}

% This section outlines the rationale behind our selection of pre-trained models, datasets, and adversarial attack approaches, emphasizing their role in enhancing both the relevance and impact of our study.

\subsubsection{Datasets}
\label{DS}
This study evaluates the quality of LLM-generated data for fine-tuning robust PTMs, comparing it to human-written data in software analytics tasks. To facilitate this evaluation, we carefully select software analytics tasks where human-written and LLM-generated data are available to solve the same task. For example, Alam et al. \cite{alam2023gptclonebench, roy2023unveiling} demonstrated that ChatGPT is capable of generating equivalent clone pairs for answering the same Stack Overflow questions from the SemanticCloneBench dataset \cite{al2020semanticclonebench}, which is called GPTCloneBench. Additionally, to ensure the generalizability of our study findings, we consider diverse tasks in software analytics. Consequently, we select three types of software analytics tasks: clone detection, code summarization, and sentiment analysis in code review discussions.

% This careful consideration of diverse tasks makes our study thorough and well-rounded.

\textbf{Code Summarization}. It is a task aimed at generating descriptions that explain the functionality of a given code snippet \cite{ahmed2024automatic}. CodeSearchNet \cite{husain2019codesearchnet} is a widely used code summarization dataset. In line with previous studies \cite{zeng2022extensive, lu2021codexglue, du2023extensive}, we focus solely on the Java sub-dataset, which consists of $164,923$ examples for training, $5,183$ for validation, and $10,955$ for testing. However, to the best of our knowledge, the LLM-generated code summary is unavailable for the Java sub-dataset. Ahmed et al. \cite{ahmed2024automatic} proposed a novel prompt engineering technique: \textit{Automatic Semantic Augmentation of Prompts (ASAP)} to generate code summaries for the proportion of the CodeSearchNet dataset. They uniformly selected $1,000$ samples from the Java sub-dataset and generated code summaries using Text-Davinci-003 \& GPT-3.5-turbo models. Since the code summary generation is quite expensive and time-consuming using \textit{ASAP}, we randomly select $10,000$ training, $2,000$ valid, and $2,000$ test samples for summary generation from the train, valid, and test samples of the Java sub-dataset, respectively.

% Our study considers \textit{ASAP} to generate summaries for the Java sub-dataset using the GPT-3.5-turbo model.

After generating code summaries, it is crucial to validate their quality and relevance to the corresponding code snippets. Recently, Mastropaolo et al. \cite{mastropaolo2024evaluating} proposed a new metric: \textit{\textbf{S}ummary al\textbf{I}gnment to co\textbf{D}e s\textbf{E}mantics (SIDE)}, to automatically measure the degree to which a generated summary aligns with the semantics of a given code snippet using contrastive learning. Unlike traditional metrics such as \textit{BLEU}, \textit{ROUGE}, and \textit{METEOR}, which measure the similarity between the reference summary and the generated summary, \textit{SIDE} provides a suitability score (A value approaching $1.0$ indicates an ideally suitable summary) indicating how relevant the generated summary is to the given code snippet, independently from the reference summary. However, \textit{SIDE} does not specify a threshold value to determine whether a generated summary is suitable; instead, it provides a score. Therefore, we set $0.5$ as the threshold to filter out irrelevant summaries from the training, validation, and test samples. After validating the generated summaries using \textit{SIDE}, we identified irrelevant summaries in $146$ training, $41$ validation, and $44$ test samples out of $10,000$ training, $2,000$ validation, and $2,000$ test samples.

To further validate the remaining irrelevant summaries filtered by the \textit{SIDE}, we engaged two experienced Java developers to review and classify them manually. Each developer independently evaluated whether the filtered summaries were relevant to their corresponding code snippets, resulting in an initial Cohen's Kappa agreement score of 0.75. As this score was considered insufficient, the developers discussed their disagreements and collaboratively reached a consensus. After resolving these disagreements, Cohen's Kappa score improved to 0.92, indicating a high level of agreement between the two developers.

\textbf{Clone Detection}. It focuses on determining whether two code snippets are identical or similar \cite{al2020semanticclonebench}. For this task, we select SemanticCloneBench \cite{al2020semanticclonebench} and GPTCloneBench \cite{alam2023gptclonebench}, as both datasets contain solutions written by humans and generated by ChatGPT for the same Stack Overflow questions. Alam et al. \cite{alam2023gptclonebench} stated that GPTCloneBench is 15 times larger than SemanticCloneBench. Therefore, to ensure a fair comparison (we maintain an almost equal sample size for training, validation, and testing and assess the fine-tuned model’s performance both before and after it undergoes adversarial attacks across all three tasks selected for this study), we preprocess GPTCloneBench so that the number of examples in both datasets is almost equal. Before starting the preprocessing on GPTCloneBench, we briefly describe how GPTCloneBench was generated by leveraging SemanticCloneBench. SemanticCloneBench has 1000 true semantic clone pairs for the Java programming language. For example, \texttt{\{$S_1$,$S_2$\}} is a clone pair. First, the clone pair is separated, and then either \texttt{$S_1$} or \texttt{$S_2$} is passed to the GPT-3 model \cite{alam2023gptclonebench}. GPTCloneBench used the prompt ``\textit{Give me 10 distinctive implementations for the following code $<$ code fragment $>$}".

It is important to note that the first code fragment in each clone pair of GPTCloneBench contains either the first or second code fragment of each SemanticCloneBench clone pair. Therefore, the GPTCloneBench dataset does not only contain the semantic equivalent code generated by the GPT-3 model of each clone pair in SemanticCloneBench. To ensure a truly semantic clone pair dataset, we further preprocess GPTCloneBench to include only clone pairs generated by the GPT-3 model, as described below: let us consider, the first code (e.g., \texttt{$S_1$}) fragment of SemanticCloneBench clone pair \texttt{\{$S_1$,$S_2$\}} has been used to generate 10 clone pairs using GPT-3: \texttt{\{$S_1$,$G_1$\}, \{$S_1$,$G_2$\}, \{$S_1$,$G_3$\}, \{$S_1$, $G_4$\}, \{$S_1$, $G_5$\}, \{$S_1$,$G_6$\}, \\ \{$S_1$,$G_7$\}, \{$S_1$,$G_8$\}, \{$S_1$,$G_9$\}, \{$S_1$,$G_{10}$\}}. We can extract the second code fragment of each clone pair and make a set of code fragments generated by GPT-3 models: \texttt{\{$G_1$,$G_2$,$G_3$,$G_4$,$G_5$,$G_6$,$G_7$, \\ $G_8$,$G_9$,$G_{10}$\}}. From this set, we can create $N$ clone pairs using the formula \(\binom{N}{k} = \frac{N!}{k!(N-k)!}\). Therefore, the total number of combinations for the given set of 10 code fragments generated by GPT-3.5 is \(\binom{10}{2} = \frac{10!}{2!(10-2)!} = 45\). 

% Our primary focus is on assessing the quality of code written by humans and generated by LLMs for robust model training under adversarial attacks in clone detection tasks.

% \[
%  \binom{10}{2} = \frac{10!}{2!(10-2)!} = 45
% \]

Here, for a fair comparison, we randomly select only one clone pair (e.g., \texttt{\{$G_1$,$G_2$\}}) out of a maximum of 45 combinations. Thus, we filter out code fragments from GPTCloneBench and create a dataset containing truly semantic clone pairs generated by the GPT-3 model. It is important to note that GPTCloneBench does not contain corresponding clone pairs for each of the clone pairs in SemanticCloneBench, as sometimes the GPT-3 model fails to generate a semantic clone \cite{alam2023gptclonebench}. Thus, following the data preprocessing steps outlined above, we finally obtain 948 truly semantic clone pairs generated by the GPT-3 model for the Java programming language. We further augment this dataset by creating combinations of semantic clones and non-clones, following the approach outlined in Arshad et al. \cite{arshad2022codebert}. Table \ref{GCB Augment} represents the augmented dataset for two clone pairs in the GPTCloneBench dataset. 

% Thus, we augment the SemanticCloneBench and GPTCloneBench datasets by six times the number of clone pairs in each dataset. For example, in the case of the GPTCloneBench dataset, we have $948 * 2 = 1896$ Type-1 clone pairs and 948 semantic clone pairs. In addition, by creating combinations, we generate $948 * 3 = 2844$ non-clone pairs. Thus, the final preprocessed SemanticCloneBench and GPTCloneBench datasets contain 6000 and 5688 clone and non-clone pairs, respectively.

\begin{table}[htbp]
\caption{Data augmentation for clone pairs in GPTCloneBench dataset.}
\vspace{-1.05em}
\centering
\resizebox{\columnwidth}{!}{
\begin{tabular}{c|c}
\hline
\{$G_1$, $G_2$\}                                                                                                                                          & \{$G_3$, $G_4$\}                                                                                                                           \\ \hline
\begin{tabular}[c]{@{}l@{}}Type-1 clone: \{$G_1$, $G_1$\}, \{$G_2$, $G_2$\}\\ Type-4 clone: \{$G_1$, $G_2$\}\end{tabular} & \begin{tabular}[c]{@{}l@{}}Not Clone: \{$G_1$, $G_3$\}, \{$G_1$, $G_4$\},\\  \{$G_2$, $G_3$\}\end{tabular} \\ \hline
\end{tabular}}
\label{GCB Augment}
\vspace{-1.25em}
\end{table}

\textbf{Sentiment Analysis}. It determines whether the comments in code review discussions are civic or non-civic \cite{sarker2023automated, rahman2024words}. For this task, we consider the incivility comment detection dataset used in developing the SOTA toxicity detector, \textit{ToxiCR} \cite{sarker2023automated}. This dataset contains only human-written review comments from open-source code review discussions. Rahman et al. \cite{rahman2024words} further augmented this dataset, generating a total of $4485$ civic and non-civic comments using ChatGPT to enhance toxicity detection. Our manual investigation reveals that human-written comments are typically longer than those generated by ChatGPT. Therefore, for a fair comparison when measuring the quality of data written by humans versus that generated by LLMs, we scale down the volume of human-written comments accordingly. We first calculate the maximum length of a single ChatGPT-generated comment and filter out human-written comments with lengths less than or equal to this maximum. Initially, we identify approximately 10k human-written comments. We randomly select $4485$ comments from this subset to fine-tune PTMs for sentiment analysis in code review discussions.

\subsubsection{Target Models}
\label{TM}
The rationale for selecting the datasets in our study is detailed in Section \ref{DS}. In the literature, numerous SOTA DL-based tools and techniques have been proposed for detecting code clones \cite{nafi2019clcdsa}, performing sentiment analysis \cite{sarker2023automated}, and generating code summaries \cite{ahmed2024automatic}. However, studies indicate that pre-trained models deliver performance comparable to SOTA approaches for these tasks \cite{feng2020codebert, rahman2024words, ahmad2021unified}. Additionally, fine-tuning PTMs for these tasks is computationally more efficient than training SOTA DL models from scratch on the same datasets \cite{devlin2018bert}. Therefore, in this study, we select PTMs for fine-tuning on the chosen datasets to assess the quality of data written by humans and generated by LLMs in the context of robust model training.

Typically, PTM is categorized into three types: encoder-only, decoder-only, and encoder-decoder models \cite{zeng2022extensive}, based on the architectures derived from the transformer model \cite{vaswani2017attention}. CodeBERT \cite{feng2020codebert} is a widely used encoder-only PTM for program understanding (e.g., clone detection \cite{nafi2019clcdsa, alam2023gptclonebench}) and generation (e.g., code summarization) tasks \cite{ahmed2024automatic, mastropaolo2024evaluating}. In sentiment analysis (e.g., classification of code review comments in discussions), encoder-only PTMs such as BERT \cite{devlin2018bert}, RoBERTa \cite{liu2019roberta}, and DistilBERT \cite{sanh2019distilbert} have been extensively used \cite{rahman2024words, sarker2023automated}. The decoder-only PTM, such as CodeGPT \cite{lu2021codexglue}, is effective for program understanding and generation tasks. Finally, encoder-decoder PTMs like PLBART \cite{ahmad2021unified} were proposed to address both code understanding and generation tasks.

Since no single PTM outperforms all other PTMs across program understanding and generation tasks, we consider one PTM for each task as the previous reports \cite{zeng2022extensive, du2023extensive}. For example, CodeT5 outperforms PLBART in vulnerability detection, while PLBART excels in clone detection. In this study, we specifically select CodeBERT, CodeGPT, and PLBART, with each model representing a distinct category. Furthermore, we choose popular encoder-only models like BERT, RoBERTa, and DistilBERT for sentiment analysis of comments in code review discussions due to their outstanding performance in sentiment analysis tasks \cite{sarker2023automated}.

% We present the SOTA PTMCs for program understanding and generation to date in Section \ref{PTMC}. Since GPTCloneBench was created by leveraging SemanticCloneBench and OpenAI's GPT-3 \cite{brown2020language} model, we therefore consider CodeGPT \cite{lu2021codexglue} as a target PTMC for our empirical investigation. Additionally, research shows that CodeBERT is highly efficient compared to other PTMCs (e.g., GraphCodeBERT, CodeT5) for code understanding tasks \cite{du2023extensive, xiao2023empirical}. As code clone detection is a code understanding task, we consider CodeBERT to be another target PTMC for our empirical study. Finally, previous studies extensively utilized CodeBERT and CodeGPT to analyze the robustness of PTMCs under adversarial attacks \cite{zeng2022extensive, du2023extensive}. Therefore, we select them as our study's target models for adversarial attacks.

\subsection{Adversarial Attack Approaches}
\label{Adv Attack}
In this empirical study, adversarial attacks play a crucial role in comparing the robustness of PTMs fine-tuned on human-written and LLM-generated data. Du et al. \cite{du2023extensive} discussed the rationale for selecting black-box attack approaches over white-box ones. Therefore, this study focuses on black-box approaches for conducting adversarial attacks on PTMs across different software analytics tasks. For code clone detection and code summarization, we consider SOTA approaches such as ALERT \cite{yang2022natural}, WIR-Random \cite{zeng2022extensive}, Metropolis-Hastings Modiﬁer (MHM) \cite{zhang2020generating}, and BeamAttack \cite{du2023extensive}. All four of these black-box adversarial attack techniques focus exclusively on identifier substitution in source code when generating adversarial examples. For sentiment analysis in code review discussions, we select TextFooler \cite{jin2020bert}, BAEAttack \cite{garg2020bae}, TextBugger \cite{li2018textbugger}, PWWSAttack \cite{ren2019generating}, and BERT-Attack \cite{li2020bert}. Each of these attack methods is designed to exploit different vulnerabilities in text models by generating semantically similar adversarial examples through the substitution of important words.

\subsection{Attack Settings}
\label{AS}
Adversarial attacks are crucial for assessing the quality of human-written and LLM-generated data in robust model training for our study. Since we conduct an empirical study, we adapt the original repository of the study's selected four attack approaches for code clone detection and summarization tasks. To do so, we leave the hyperparameters of these attack approaches unaltered. Additionally, we employ the TextAttack \cite{morris2020textattack} library to carry out five selected adversarial attacks on PTMs fine-tuned for sentiment analysis in code review comments. Since the number of sub-word combinations grows exponentially for BERT-Attack\footnote{https://github.com/QData/TextAttack/issues/586}, we set the maximum number of sub-word candidates to $8$. We also retain the default hyperparameter settings for the other four attack approaches for the sentiment analysis of code review comments. Section \ref{DS} outlines the preprocessing applied to the selected datasets to ensure a fair evaluation of the quality of human-written and LLM-generated data for robust model training. For model fine-tuning on sentiment analysis, we employ the split ratio of 70\%-15\%-15\%, where 70\% of the data is used for fine-tuning, 15\% for validation, and 15\% for testing \cite{rahman2024words, sarker2023automated}. For the other two tasks, we sample the training, validation, and testing data following prior works \cite{du2023extensive, zeng2022extensive, yang2022natural}. Since the four attack approaches for code clone detection and summarization tasks only involve identifier renaming while generating adversarial examples, we exclude test code snippets without identifiers. Finally, we consider only instances correctly predicted by PTMs for conducting adversarial attacks, a common practice in existing studies \cite{zhang2020generating, yang2022natural}.

\subsection{Evaluation Metrics}
\label{EM}
% We utilize two sets of evaluation metrics in our experiment. The first set compares the performance of PTMs fine-tuned on human-written and LLM-generated data. The second set evaluates the robustness of PTMs when subjected to adversarial attacks. 

% This evaluation aims to assess the quality of both human-written and LLM-generated data for robust model training.

\subsubsection{Evaluating Performance of PTMs}
\label{Eve PTMs}
We utilize the traditional metrics to evaluate the performance of PTMs. \textbf{Accuracy.} It represents the ratio of correctly predicted instances to the total number of instances in the test set. \textbf{Precision (P).} It is the ratio of true positive predictions to the total number of instances predicted as positive. \textbf{Recall (R).} It is the ratio of true positive predictions to the total number of actual positive instances. \textbf{F1 Score.} It is calculated as the harmonic mean of precision and recall: $F_1 = 2 * (P * R) / (P + R)$. \textbf{BLEU-4.} \cite{zeng2022extensive} is a variation of the BLEU metric that is commonly used to evaluate the similarity between generated text and the actual ground truth. 

% The $4$ in BLEU-4 refers to the use of four consecutive words, known as 4-grams, as the matching unit for comparison.

\subsubsection{Assessing Robustness of PTMs}
\label{Ase PTMs}
To evaluate the robustness of PTMs after fine-tuning on the human-written and LLM-generated data for the selected tasks in our study, we perform both quantitative and qualitative analyses.

\textbf{Quantitative.} For the quantitative analysis, we use \textit{Attack Success Rate} (\%ASR) \cite{yang2022natural} and \textit{Average Model Query} (AMQ) \cite{du2023extensive} metrics. \textbf{\%ASR} represents the percentage of test samples where an attack approach successfully generates adversarial examples. A higher \%ASR signifies that a PTM is less robust against such attacks. \textbf{AMQ} denotes the number of queries directed at the targeted model when generating adversarial examples. A higher AMQ value indicates greater model robustness against adversarial attacks.

\textbf{Qualitative.} We conduct a qualitative analysis to assess the quality of the generated adversarial examples, which are essential for adversarial fine-tuning to improve model robustness \cite{yang2022natural}. Since the inputs to the PTMs for code clone detection and summarization tasks are code, we adopt the existing metrics: \textit{Identifier Change Rate} (ICR), \textit{Token Change Rate} (TCR), \textit{Average Edit Distance} (AED) and \textit{Average Code Similarity} (ACS) from Du et al. \cite{du2023extensive} to assess the quality of the generated adversarial examples. In contrast, the inputs to the PTMs for sentiment analysis consist of natural text. For this purpose, we apply different similarity measures, including the Sentence-BERT (SBERT) \cite{reimers2019sentence}, \textit{Jaccard Similarity Coefficient} (JSC), \textit{Dice Similarity Coefficient} (DSC), \textit{Levenshtein Distance} (LD), \textit{Levenshtein Ratio} (LR), \textit{Jaro Similarity} (JS), and \textit{Jaro-Winkler similarity} (JWS), which are commonly used in the fields of information retrieval and data mining. 

\textbf{ICR.} For a set of $m$ adversarial examples, if there are a total of $k_i$ identifiers in the $i-$th code snippet and $n_i$ identifiers have been altered in generating adversarial examples, the ICR is calculated as $\sum_{1}^{m} n_i / \sum_{1}^{m} k_i$. \textbf{TCR.} It represents the ratio of modified tokens in the adversarial example to the total number of tokens in the code. \textbf{ACS.} It is determined using cosine similarity, which is based on the computed embeddings from CodeBERT before and after carrying out adversarial attacks. \textbf{AED.} It quantifies the character-level token variances, indicating how many edits are required at the character level to change one token into another. \textit{Typically, a high-quality adversarial example should have lower ICR, AED, and TCR values while displaying a higher ACS}.

\textbf{SBERT}. Using a Siamese network structure, SBERT improves upon BERT by producing effective and high-quality sentence embeddings suitable for computing semantic similarity. A higher similarity score indicates a greater degree of semantic resemblance. \textbf{JSC} and \textbf{DSC} measure the overlap between sets, with both metrics ranging from 0 (no overlap) to 1 (identical). JSC is defined as $JSC(A, B) = \frac{|A \cap B|}{|A \cup B|}$, while DSC is defined as $DSC(A, B) = \frac{2 * |A \cap B|}{|A| + |B|}$. \textbf{LD} quantifies string dissimilarity by counting the minimum character edits (e.g., insertions, deletions, or substitutions) needed for transformation. In contrast, \textbf{LR} normalizes this distance to a $0-1$ similarity score. Finally, \textbf{JS} and \textbf{JWS} measure string similarity based on matching characters and transpositions, with higher values indicating greater similarity. Following \cite{du2023extensive}, we report average similarity values to compare the quality of adversarial examples generated from human-written and LLM-generated data.

\subsection{Research Questions}
\label{RQ}
This study compares the quality of data generated by LLMs with human-written data for fine-tuning PTMs in software analytics tasks. We assess data quality under adversarial attacks, focusing on the robustness of the fine-tuned PTMs. We also investigate the quality of generated adversarial examples, as prior research highlights its significant impact on the effectiveness of adversarial fine-tuning \cite{yang2022natural}. Thus, We formulate our target RQs as follows:

% \textbf{RQ1: (Model performance)} When fine-tuned on human-written and LLM-generated data, how well do pre-trained models perform in software analytics tasks? 

\textbf{RQ1: (Model's performance)} How well do pre-trained models perform in software analytics tasks when fine-tuned on human-written and LLM-generated data? In this RQ, we thoroughly compare the performance of PTMs fine-tuned on the human-written and LLM-generated data using the metrics outlined in \ref{Eve PTMs}. \textbf{RQ2: (Model's Robustness)} To what extent does the performance of fine-tuned pre-trained models decline under adversarial attacks in software analytics tasks? This RQ quantitatively compares the performance decline of PTMs fine-tuned on human-written and LLM-generated data using the \%ASR and AMQ metrics outlined in Section \ref{Ase PTMs}. \textbf{RQ3: (Adversarial examples quality)} What is the quality of adversarial examples generated by attacks on fine-tuned pre-trained models? This research question qualitatively assesses the quality of the adversarial examples. This assessment involves utilizing eleven similarity measure metrics outlined in Section \ref{Ase PTMs}.

% How do pre-trained models fine-tuned on human-written versus LLM-generated data perform in software analytics tasks?

% This research question qualitatively assesses the quality (e.g., Naturalness\footnote{Naturalness signifies that the adversarial examples should be as similar as possible with the original one. }) of the adversarial examples. This assessment involves utilizing in total ten similarity measure metrics outlined in Section \ref{Ase PTMs}.

% Naturalness plays a key role in generating adversarial examples, as emphasized by \cite{zhou2022adversarial, yang2022natural}.

% This research question qualitatively assesses the quality of the adversarial examples the performance decline of PTMs fine-tuned on human-written and LLM-generated data using the ASR and AMQ metrics outlined in Section \ref{Ase PTMs}.

\section{Empirical Results}
\label{exp_res}
This section presents the results of our extensive experiments. Our study consists of 39 experimental combinations: $3 \text{ models}*5 \text{ attacks}=15$ for sentiment analysis, $3 \text{ models}*4 \text{ attacks}=12$ for clone detection, and $3\text{ models}*4 \text{ attacks}=12$ for code summarization tasks. 

% We implement and evaluate our experiments using the PyTorch library\footnote{https://pytorch.org/}. Our study involves two datasets, four SOTA adversarial attack techniques, and two SOTA PTMCs. Similarly to prior studies \cite{yang2022natural, hashemi2020permuteattack, ballet2019imperceptible, zhang2020generating}, our study concentrates solely on the percentage of accurately predicted instances in the test dataset during adversarial attacks. Thus, we first analyze the performance of CodeBERT and CodeGPT fine-tuned on the SemanticCloneBench and GPTCloneBench datasets, and then, we present the results of our empirical investigation by addressing the research question outlined in Section \ref{method}.

% \begin{table}[ht]
% \caption{Comparison of performance between HUM and LLM models.}
%     \centering
%     \begin{tabular}{|c|c|c|}
%         \hline
%         Metric & HUM & LLM \\
%         \hline
%         Accuracy & 0.94 & 0.96 \\
%         Precision & 0.94 & 0.96 \\
%         Recall & 0.94 & 0.96 \\
%         \hline
%     \end{tabular}
%     \centering
%     \caption* {\threeDcircle{green} indicates the higher value, \threeDcircle{red} indicates the lower value.}
% \end{table}

\subsection{Evaluating Performance of PTMs}
\label{PTMs perform}
The performance of pre-trained models fine-tuned on human-written and LLM-generated data prior to adversarial attacks is presented in Table \ref{RQ1Results}. The results highlight the differences\footnote{Diff represents the difference in metric values between the performance of models fine-tuned on human-written data and LLM-generated data. A positive Diff indicates that the model performs better on human-written data, while a negative Diff indicates better performance on LLM-generated data.} in model performance based on accuracy (Acc), precision (Prec), recall (Rec), and F1-score (F1-s). For the sentiment analysis task, all models demonstrate a slight improvement across all the metrics when fine-tuned on LLM-generated data, except in two cases. For instance, the accuracy, precision, recall, and F1-score increased by 0.02 for the BERT model, suggesting a modest enhancement in performance when using LLM-generated data. Recall demonstrates more variation, with RoBERTa experiencing a notable improvement from 0.94 to 0.98. BERT and DistilBERT saw increases of 0.94 to 0.96 and 0.95 to 0.97, respectively. In contrast, the precision and F1-score show minor improvements, including two cases where the performances are equal. Overall, the results suggest that LLM-generated data slightly improves the performance of PTMs over human-written data for the sentiment analysis task. We observe similar results in the code summarization task as presented in Table \ref{RQ1Results}, where PTMs fine-tuned on LLM-generated data perform slightly better than those fine-tuned on human-written data, as indicated by the BLEU-4 scores.

%For precision, LLM-generated data also exhibit slight improvements, with BERT increasing from 0.94 to 0.96, while DistilBERT remained unchanged at 0.95. These results highlight a slight edge for LLM-generated data in identifying true positives. BERT and DistilBERT saw smaller increases (0.94 to 0.96 and 0.95 to 0.97, respectively), indicating that LLM-generated data helps capture more relevant sentiment instances. Similar to other metrics, the F1-score shows minor improvements for all models when fine-tuned on LLM-generated data, with differences ranging from 0.01 to 0.02.

\begin{table}[htbp]
\centering
\caption{Performance comparison of PTMs fine-tuned on different software analytics tasks.}
\vspace{-1.05em}
\resizebox{\columnwidth}{!}{ % Adjusts the table to fit within the column width
\begin{tabular}{llllllllll|}
\hline
\multicolumn{1}{|l|}{\multirow{3}{*}{Metric}} & \multicolumn{9}{c|}{\textbf{Sentiment Analysis}}                                                                                                                                                                                                                  \\ \cline{2-10} 
\multicolumn{1}{|l|}{}                        & \multicolumn{3}{c|}{\textbf{BERT}}                                                   & \multicolumn{3}{c|}{\textbf{RoBERTa}}                                                & \multicolumn{3}{c|}{\textbf{DistilBERT}}                                            \\ \cline{2-10} 
\multicolumn{1}{|l|}{}                        & \multicolumn{1}{c|}{HUM}   & \multicolumn{1}{c|}{LLM}   & \multicolumn{1}{c|}{Diff}  & \multicolumn{1}{c|}{HUM}   & \multicolumn{1}{c|}{LLM}   & \multicolumn{1}{c|}{Diff}  & \multicolumn{1}{c|}{HUM}   & \multicolumn{1}{c|}{LLM}   & \multicolumn{1}{c|}{Diff} \\ \hline
\multicolumn{1}{|l|}{Accuracy}                & \multicolumn{1}{l|}{0.94 \threeDcircle{red}}  & \multicolumn{1}{l|}{0.96 \threeDcircle{green}}  & \multicolumn{1}{l|}{-0.02} & \multicolumn{1}{l|}{0.95 \threeDcircle{red}}  & \multicolumn{1}{l|}{0.96 \threeDcircle{green}}  & \multicolumn{1}{l|}{-0.01} & \multicolumn{1}{l|}{0.95 \threeDcircle{red}}  & \multicolumn{1}{l|}{0.96 \threeDcircle{green}}  & -0.01                     \\ \hline
\multicolumn{1}{|l|}{Precision}               & \multicolumn{1}{l|}{0.94 \threeDcircle{red}}  & \multicolumn{1}{l|}{0.96 \threeDcircle{green}}  & \multicolumn{1}{l|}{-0.02} & \multicolumn{1}{l|}{0.96 \threeDcircle{green}}  & \multicolumn{1}{l|}{0.95 \threeDcircle{red}}  & \multicolumn{1}{l|}{+0.01} & \multicolumn{1}{l|}{0.95 \threeDcircle{yellow}}  & \multicolumn{1}{l|}{0.95 \threeDcircle{yellow}}  & 0.00                      \\ \hline
\multicolumn{1}{|l|}{Recall}                  & \multicolumn{1}{l|}{0.94 \threeDcircle{red}}  & \multicolumn{1}{l|}{0.96 \threeDcircle{green}}  & \multicolumn{1}{l|}{-0.02} & \multicolumn{1}{l|}{0.94 \threeDcircle{red}}  & \multicolumn{1}{l|}{0.98 \threeDcircle{green}}  & \multicolumn{1}{l|}{-0.04} & \multicolumn{1}{l|}{0.95 \threeDcircle{red}}  & \multicolumn{1}{l|}{0.97 \threeDcircle{green}}  & -0.02                     \\ \hline
\multicolumn{1}{|l|}{F1-Score}                & \multicolumn{1}{l|}{0.94 \threeDcircle{red}}  & \multicolumn{1}{l|}{0.96 \threeDcircle{green}}  & \multicolumn{1}{l|}{-0.02} & \multicolumn{1}{l|}{0.95 \threeDcircle{red}}  & \multicolumn{1}{l|}{0.96 \threeDcircle{green}}  & \multicolumn{1}{l|}{-0.01} & \multicolumn{1}{l|}{0.95 \threeDcircle{red}}  & \multicolumn{1}{l|}{0.96 \threeDcircle{green}}  & -0.01                     \\ \hline
\multicolumn{10}{|c|}{\textbf{Clone Detection}}                                                                                                                                                                                                                                                                   \\ \hline
\multicolumn{1}{|l|}{\multirow{2}{*}{Metric}} & \multicolumn{3}{c|}{\textbf{CodeBERT}}                                               & \multicolumn{3}{c|}{\textbf{CodeGPT}}                                                & \multicolumn{3}{c|}{\textbf{PLBART}}                                                \\ \cline{2-10} 
\multicolumn{1}{|l|}{}                        & \multicolumn{1}{c|}{HUM}   & \multicolumn{1}{c|}{LLM}   & \multicolumn{1}{c|}{Diff}  & \multicolumn{1}{c|}{HUM}   & \multicolumn{1}{c|}{LLM}   & \multicolumn{1}{c|}{Diff}  & \multicolumn{1}{c|}{HUM}   & \multicolumn{1}{c|}{LLM}   & \multicolumn{1}{c|}{Diff} \\ \hline
\multicolumn{1}{|l|}{Accuracy}                & \multicolumn{1}{l|}{0.52 \threeDcircle{green}}  & \multicolumn{1}{l|}{0.39 \threeDcircle{red}}  & \multicolumn{1}{l|}{+0.13} & \multicolumn{1}{l|}{0.64 \threeDcircle{green}}  & \multicolumn{1}{l|}{0.57 \threeDcircle{red}}  & \multicolumn{1}{l|}{+0.07} & \multicolumn{1}{l|}{0.81 \threeDcircle{green}}  & \multicolumn{1}{l|}{0.69 \threeDcircle{red}}  & +0.12                     \\ \hline
\multicolumn{1}{|l|}{Precision}               & \multicolumn{1}{l|}{0.40 \threeDcircle{green}}   & \multicolumn{1}{l|}{0.34 \threeDcircle{red}}  & \multicolumn{1}{l|}{+0.06} & \multicolumn{1}{l|}{0.63 \threeDcircle{green}}  & \multicolumn{1}{l|}{0.56 \threeDcircle{red}}  & \multicolumn{1}{l|}{+0.07} & \multicolumn{1}{l|}{0.81 \threeDcircle{green}}  & \multicolumn{1}{l|}{0.69 \threeDcircle{red}}  & +0.12                     \\ \hline
\multicolumn{1}{|l|}{Recall}                  & \multicolumn{1}{l|}{0.50 \threeDcircle{green}}   & \multicolumn{1}{l|}{0.25 \threeDcircle{red}}  & \multicolumn{1}{l|}{+0.25} & \multicolumn{1}{l|}{0.64 \threeDcircle{green}}  & \multicolumn{1}{l|}{0.57 \threeDcircle{red}}  & \multicolumn{1}{l|}{+0.07} & \multicolumn{1}{l|}{0.81 \threeDcircle{green}}  & \multicolumn{1}{l|}{0.69 \threeDcircle{red}}  & +0.12                     \\ \hline
\multicolumn{1}{|l|}{F1-Score}                & \multicolumn{1}{l|}{0.34 \threeDcircle{green}}  & \multicolumn{1}{l|}{0.29 \threeDcircle{red}}  & \multicolumn{1}{l|}{+0.05} & \multicolumn{1}{l|}{0.63 \threeDcircle{green}}  & \multicolumn{1}{l|}{0.57 \threeDcircle{red}}  & \multicolumn{1}{l|}{+0.06} & \multicolumn{1}{l|}{0.81 \threeDcircle{green}}  & \multicolumn{1}{l|}{0.69 \threeDcircle{red}}  & +0.12                     \\ \hline
\multicolumn{10}{|c|}{\textbf{Code Summarization}}                                                                                                                                                                                                                                                                \\ \hline
\multicolumn{1}{|l|}{\multirow{2}{*}{Metric}} & \multicolumn{3}{c|}{\textbf{CodeBERT}}                                               & \multicolumn{3}{c|}{\textbf{CodeGPT}}                                                & \multicolumn{3}{c|}{\textbf{PLBART}}                                                \\ \cline{2-10} 
\multicolumn{1}{|l|}{}                        & \multicolumn{1}{c|}{HUM}   & \multicolumn{1}{c|}{LLM}   & \multicolumn{1}{c|}{Diff}  & \multicolumn{1}{c|}{HUM}   & \multicolumn{1}{c|}{LLM}   & \multicolumn{1}{c|}{Diff}  & \multicolumn{1}{c|}{HUM}   & \multicolumn{1}{c|}{LLM}   & \multicolumn{1}{c|}{Diff} \\ \hline
\multicolumn{1}{|l|}{BLEU-4}                  & \multicolumn{1}{l|}{13.95 \threeDcircle{red}} & \multicolumn{1}{l|}{18.65 \threeDcircle{green}} & \multicolumn{1}{l|}{-4.70} & \multicolumn{1}{l|}{13.39 \threeDcircle{red}} & \multicolumn{1}{l|}{15.06 \threeDcircle{green}} & \multicolumn{1}{l|}{-1.67} & \multicolumn{1}{l|}{12.09 \threeDcircle{red}} & \multicolumn{1}{l|}{16.01 \threeDcircle{green}} & -3.92                     \\ \hline
\end{tabular}
}
\caption* {\footnotesize \textnormal {\threeDcircle{green} indicates the higher value, \threeDcircle{red} indicates the lower value, \threeDcircle{yellow} indicates no difference}}
\label{RQ1Results}
\vspace{-1.5em}
\end{table}

In the experimental results for clone detection, as shown in Table \ref{RQ1Results}, all the PTMs consistently perform better when fine-tuned on human-written data compared to LLM-generated data across all metrics. Specifically, CodeBERT demonstrates a higher performance in accuracy, precision, recall, and F1-score when fine-tuned on human-written data. For example, its accuracy increases from 0.39 to 0.52, representing an improvement of +0.13. Similarly, recall improves significantly from 0.25 to 0.50, with a difference of +0.25 when fine-tuned on human-written data. These results show that PTMs fine-tuned on human-written data significantly outperform those fine-tuned on LLM-generated data.

\begin{tcolorbox}[
    enhanced,
    attach boxed title to top left={yshift=-3mm,yshifttext=-1mm}, % Attach title to top left
    colback=mycolor_box,                 % Background color of the box
    colframe=black,                % Frame color of the box
    colbacktitle= mycolor_title,            % Background color of the title box
    coltitle=black,                % Title text color
    title=Result RQ1,            % Title text
    fonttitle=\bfseries,           % Title font
    boxed title style={size=small},% Title box style
    width=0.49\textwidth,            % Set the box width to 50% of the text width
    boxsep=1mm,                    % Reduce padding inside the box
    left=0mm,                      % Remove left padding
    right=0mm, % Remove right padding
    bottom=0mm,
]
Pre-trained models fine-tuned on LLM-generated data perform slightly better than those on human-written data in sentiment analysis and code summarization tasks, highlighting LLMs' strength in natural text generation. In clone detection, models fine-tuned on human-written data significantly outperform those on LLM-generated data, emphasizing LLMs' weakness in code generation.
    
    % Pre-trained models fine-tuned on LLM-generated data perform slightly better than those fine-tuned on human-written data in sentiment analysis and code summarization tasks, highlighting the strength of LLMs in natural text generation. On the other hand, in the clone detection task, models fine-tuned on human-written data significantly outperform those fine-tuned on LLM-generated data, emphasizing the limitations of LLMs in code generation.
\end{tcolorbox}
\label{PTMs_result}
%\end{center}

\subsection{Evaluating Robustness of PTMs}
\label{PTMs Robustness}

\begin{figure}[htbp]
  \centering
  \subfigure[\%ASR on Sentiment Analysis]{\label{ASR_SA}\includegraphics[width=3.5cm, height=2.75cm]{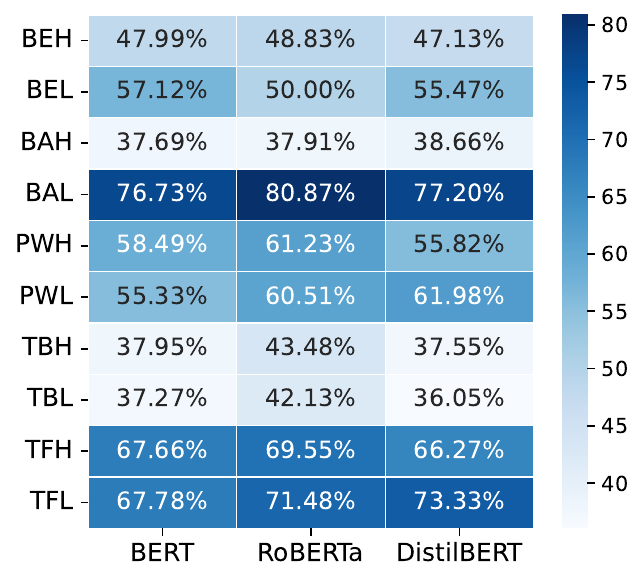}}
  \subfigure[AMQ on Sentiment Analysis]{\label{AMQ_SA}\includegraphics[width=3.5cm, height=2.75cm]{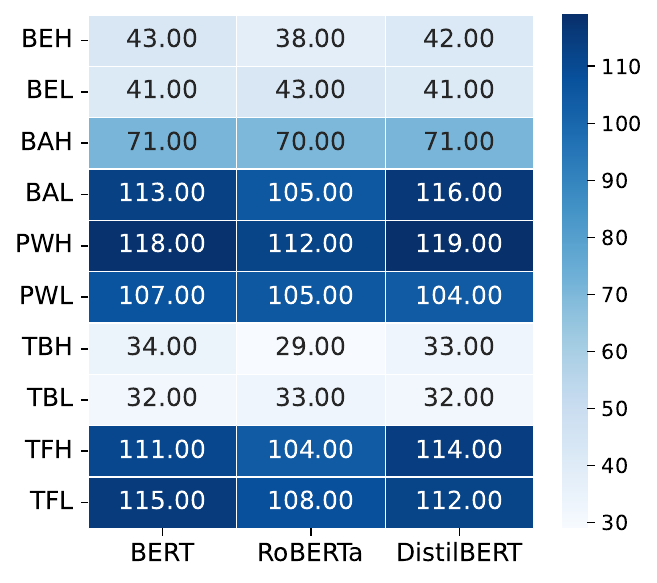}}
  \subfigure[\%ASR on Clone Detection]{\label{ASR_CD}\includegraphics[width=3.5cm, height=2.75cm]{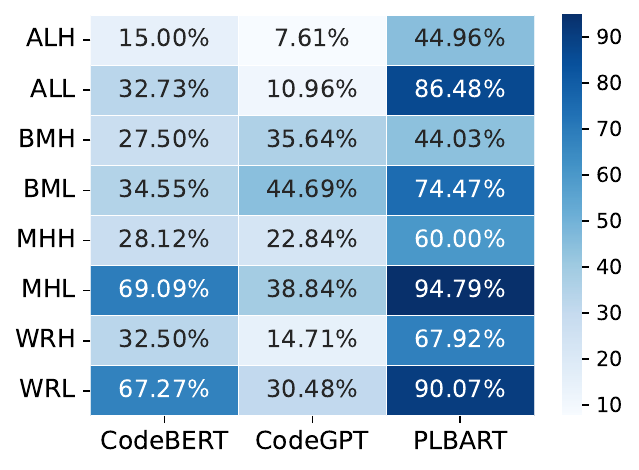}}
  \subfigure[AMQ on Clone Detection]{\label{AMQ_CD}\includegraphics[width=3.5cm, height=2.75cm]{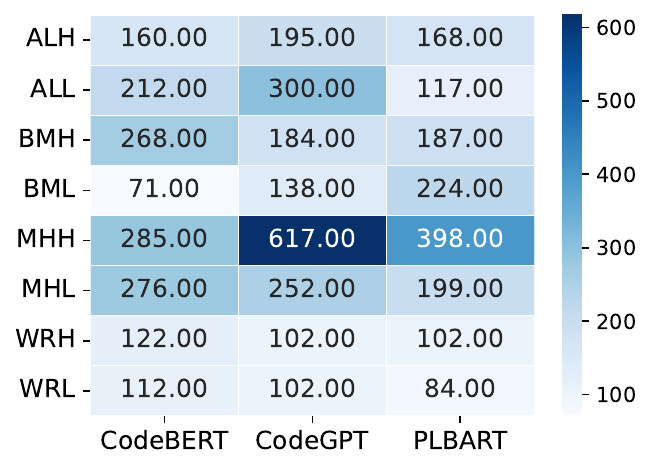}}
  \subfigure[\%ASR on Code Summarization]{\label{ASR_CS}\includegraphics[width=3.5cm, height=2.75cm]{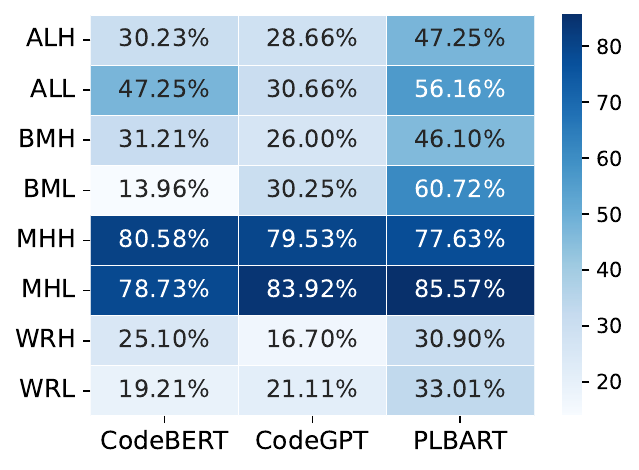}}
  \subfigure[AMQ on Code Summarization]{\label{AMQ_CS}\includegraphics[width=3.5cm, height=2.75cm]{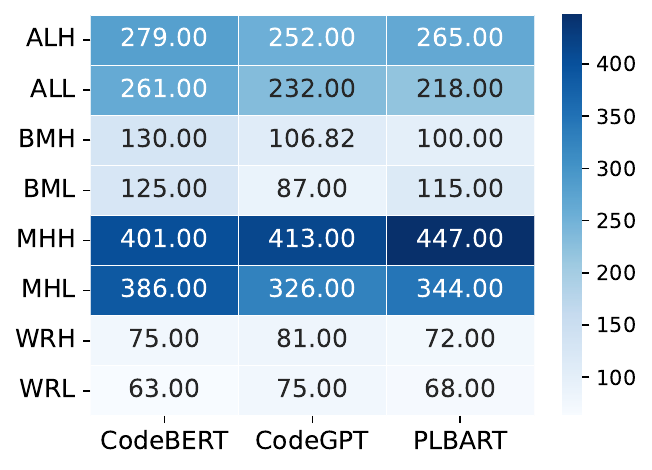}}
  \vspace{-1.05em}
  \caption{A heatmap showing \%ASR and AMQ metric values of fine-tuned models under adversarial attacks using human-written and LLM-generated data. Here, \_.\_H \& \_.\_L denote the adversarial attacks\protect\footnotemark performed on PTMs fine-tuned on human-written and LLM-generated data, respectively.}
  \label{heatmap}
  \Description{Comparing model performance under adversarial attacks.}
  \vspace{-1.75em}
\end{figure}

After evaluating the performance of the PTMs fine-tuned on human-written and LLM-generated data based on accuracy, precision, recall, and F1-score in RQ1, we now turn to evaluate their robustness under adversarial attacks using the \%ASR and AMQ metrics. The heatmaps shown in Figure \ref{heatmap} represent the \%ASR and AMQ metric values of fine-tuned models on human-written and LLM-generated data after adversarial attacks.

% After evaluating the performance of the PTMs fine-tuned on human-written and LLM-generated data based on accuracy, precision, recall, and F1-score in RQ1, we now turn to evaluate their robustness under adversarial attacks using the \%ASR and AMQ metrics. The \%ASR metric measures how successfully the adversarial attacks fool the models into changing their predictions, with higher values indicating lower model robustness against adversarial attacks. In contrast, a higher AMQ suggests that the model is more resilient to adversarial attacks. The heatmaps shown in Figure \ref{heatmap} represent the \%ASR and AMQ metric values of fine-tuned models on human-written and LLM-generated data after adversarial attacks.

% In RQ2, we assess model performance using the ASR and AMQ metrics, which reveal how well the models withstand adversarial attacks. 

\footnotetext{AL: ALERT, BM: BeamAttack, MH: Metropolis-Hasting Modifier, WR: WIR-Random, BE: BERT-Attack, BA: BAEAttack, PW: PWWSAttack, TB: TextBugger, TF: TextFooler}

A notable trend is that PTMs fine-tuned on LLM-generated data perform poorly when applying BERT-Attack, BAEAttack, and TextFooler attacks than those fine-tuned on human-written data for sentiment analysis. The difference in performance is quite significant. For instance, when applying BAEAttack to the RoBERTa model, we observe $80.87\% - 37.91\% = 42.96\%$ more performance drop in the model's prediction. For the TextBugger attack, we observe that PTMs fine-tuned on LLM-generated data show slightly better robustness than those fine-tuned on human-written data across all models, though the difference is minimal. For the sentiment analysis task, out of 15 experimental combinations, the PTMs fine-tuned on human-written data show robustness in $(10 / 15) * 100 = 67\%$ of the experimental combinations. In contrast, for LLM-generated data, the robustness is observed in only $(5 / 15) * 100 = 33\%$ of the experimental combinations. Similarly, in the code summarization task, as shown in Figure \ref{ASR_CS}, the PTMs fine-tuned on human-written data demonstrate greater robustness than those fine-tuned on LLM-generated data across 9 out of 12 (e.g., $(9/12)*100=75\%$) experimental settings. In contrast, for LLM-generated data, the robustness is observed in only $(3 / 12) * 100 = 25\%$ of the experimental combinations. In the clone detection task, as illustrated in Figure \ref{ASR_CD}, we observe that PTMs fine-tuned on human-written data exhibit more robustness than those fine-tuned on LLM-generated data across all adversarial attacks and experimental combinations. In some cases, the difference in the \%ASR metric is quite significant. 

% For example, regarding the ALERT attack on the PLBART model fine-tuned on LLM-generated data, we observe $86.48\% - 44.96\% = 41.52\%$  more performance drop in the model's predictions. 

% For instance, when applying BAEAttack to the RoBERTa model, we observe $80.87\% - 37.91\% = 42.96\%$ more performance drop in the model's prediction.

%Overall, we conclude that human-written data is superior to LLM-generated data for fine-tuning robust models.

% the performance of PTMs fine-tuned on human-written and LLM-generated data is comparable (i.e., with no clear winner) in terms of the ASR metric.

Based on the \%ASR metric, we observe that PTMs fine-tuned on human-written data exhibit greater robustness than those fine-tuned on LLM-generated data under adversarial attacks, indicating that human-written data remains a better choice for robust model training or fine-tuning. While evaluating robustness against adversarial attacks using the \%ASR metric is important, it represents only one aspect of the analysis. To provide a more comprehensive evaluation, we also assess the models' robustness using the AMQ metric, which measures the number of model queries required during adversarial attacks. By comparing the models with both metrics, we aim to gain a thorough understanding of their robustness in adversarial scenarios.

From Figures \ref{AMQ_SA}, \ref{AMQ_CD}, and \ref{AMQ_CS}, a general trend observed is that models fine-tuned on human-written data tend to exhibit higher AMQ values compared to those fine-tuned on LLM-generated data across all the selected software analytics tasks. In quantity, across all the tasks, PTMs fine-tuned on human-written data exhibit greater AMQ values for $(26 / 39) * 100 = 67\%$ experimental combinations. In addition, The difference between the AMQ values for the PTMs fine-tuned on the human-written and LLM-generated data in some cases is quite substantial. For example, While applying an MHM attack on the RoBERTa model for the clone detection task, the difference is $617 - 252 = 365$. Overall, the higher AMQ values for PTMs fine-tuned on the human-written data indicate that such data is a more practical choice for robust model training or fine-tuning in software analytics tasks than LLM-generated data.

%\begin{center}
\begin{tcolorbox}[
    enhanced,
    attach boxed title to top left={yshift=-3mm,yshifttext=-1mm}, % Attach title to top left
    colback=mycolor_box,                 % Background color of the box
    colframe=black,                % Frame color of the box
    colbacktitle= mycolor_title,            % Background color of the title box
    coltitle=black,                % Title text color
    title=Result RQ2,            % Title text
    fonttitle=\bfseries,           % Title font
    boxed title style={size=small},% Title box style
    width=0.49\textwidth,            % Set the box width to 50% of the text width
    boxsep=1mm,                    % Reduce padding inside the box
    left=0mm,                      % Remove left padding
    right=0mm, % Remove right padding
    bottom=0mm,
]

    Pre-trained models fine-tuned on human-written data are more robust than those fine-tuned on LLM-generated data under adversarial attacks, highlighting the superior quality of human-written data for developing robust models.

    % Pre-trained models fine-tuned on human-written data exhibit greater robustness than those fine-tuned on LLM-generated data under adversarial attacks, highlighting the superiority of human-written data in developing robust models.
    
    % Overall, pre-trained models fine-tuned on human-written data outperform those fine-tuned on LLM-generated data across all tasks, as indicated by the \%ASR and AMQ metrics. This finding emphasizes the effectiveness of using human-written data for robust model training or fine-tuning, particularly in the context of software analytics tasks that are subject to adversarial attacks.
    
% Pre-trained models fine-tuned on human-written data perform better than those on LLM-generated data across all the selected tasks under adversarial attacks regarding ASR and AMQ metrics, highlighting the superiority of human-written data over LLM-generated data in training or fine-tuning models for software analysis tasks.
    
    % Pre-trained models fine-tuned on LLM-generated data perform slightly better than those fine-tuned on human-written data in sentiment analysis and code summarization tasks, highlighting the strength of LLMs in natural text generation. On the other hand, in the clone detection task, models fine-tuned on human-written data significantly outperform those fine-tuned on LLM-generated data, emphasizing the limitations of LLMs in code generation.
\end{tcolorbox}
\label{PTMs_result_attack}
%\end{center}

\subsection{Evaluating Adversarial Examples Quality}
\label{PTMs Quality}

After evaluating the performance of PTMs fine-tuned on human-written and LLM-generated data before adversarial attacks (RQ1) and analyzing their robustness under adversarial conditions (RQ2), we now focus on assessing the quality of the generated adversarial examples. The results\footnote{Each cell value under each similarity measure metric contains two values separated by the `|' symbol. The first value represents the average metric values computed from PTMs fine-tuned on human-written data. In contrast, the second value corresponds to the PTMs fine-tuned on LLM-generated data. Dark grey cells denote the quality of adversarial examples is better for PTMs fine-tuned on LLM-generated data, whereas light grey cells represent a tie. This is also applicable to the other two tasks as well.} presented in Table \ref{adv_quality_cd} compare the similarity of adversarial examples to the original ones generated from PTMs fine-tuned on human-written versus LLM-generated data in the clone detection task. The comparison is based on similarity metrics, including ICR, TCR, ACS, and AED. Across most attack techniques, adversarial examples generated from PTMs fine-tuned on human-written data exhibit lower ICR values than those generated from PTMs fine-tuned on LLM-generated data. This pattern is consistent across nearly all PTMs and attack techniques and is also true for the TCR metric. 

% This assessment will compare the impact of adversarial attacks on models trained with human-written versus LLM-generated data, providing insights into the quality of the adversarial examples and highlighting how the data source—whether human-written or LLM-generated—affects model robustness. 

ACS values are relatively stable across both data, with minimal differences observed between human-written and LLM-generated data. For instance, under the BeamAttack for CodeGPT, ACS values are 0.99 for both human-written and LLM-generated data. However, AED values reveal a noticeable gap, with adversarial examples generated for LLM-generated data often exhibiting shorter edit distances. For example, under the WIR-Random attack for CodeBERT, AED values are 133 (human-written) versus 94 (LLM-generated). Overall, the experimental findings suggest that the quality of adversarial examples generated from PTMs fine-tuned on human-written data is higher than that of those generated from PTMs fine-tuned on LLM-generated data in clone detection.

% \usepackage{multirow}
\begin{table}[htbp]
\centering
\caption{Comparison of the quality of adversarial examples in the clone detection task}
\vspace{-1.05em}
\resizebox{\linewidth}{!}{ % Adjusts the table to fit within the column width
\begin{tabular}{l|llll|llll|llll}
\hline
                                  & \multicolumn{4}{c|}{\textbf{CodeBERT}}                                                                                                                                                            & \multicolumn{4}{c|}{\textbf{CodeGPT}}                                                                                                                                                             & \multicolumn{4}{c}{\textbf{PLBART}}                                                                                                                                      \\ \cline{2-13} 
\multirow{-2}{*}{\textbf{Attack}} & \multicolumn{1}{l|}{\textbf{ICR}}                          & \multicolumn{1}{l|}{\textbf{TCR}}  & \multicolumn{1}{l|}{\textbf{ACS}}                          & \textbf{AED}                       & \multicolumn{1}{l|}{\textbf{ICR}}                          & \multicolumn{1}{l|}{\textbf{TCR}} & \multicolumn{1}{l|}{\textbf{ACS}}                          & \textbf{AED}                        & \multicolumn{1}{l|}{\textbf{ICR}}  & \multicolumn{1}{l|}{\textbf{TCR}} & \multicolumn{1}{l|}{\textbf{ACS}}                          & \textbf{AED}                        \\ \hline
\textbf{AL}                       & \multicolumn{1}{l|}{\cellcolor[HTML]{C0C0C0}7.43   | 15.7} & \multicolumn{1}{l|}{1.29   | 2.1}  & \multicolumn{1}{l|}{0.99   | 0.98}                         & \cellcolor[HTML]{C0C0C0}33   | 30  & \multicolumn{1}{l|}{4.68   | 8.67}                         & \multicolumn{1}{l|}{0.63   | 1.5} & \multicolumn{1}{l|}{0.98   | 0.97}                         & 34   | 63                           & \multicolumn{1}{l|}{28.6   | 47.6} & \multicolumn{1}{l|}{4.6   | 7.21} & \multicolumn{1}{l|}{\cellcolor[HTML]{EFEFEF}0.99   | 0.99} & \cellcolor[HTML]{C0C0C0}55   | 42   \\ \hline
\textbf{BM}                       & \multicolumn{1}{l|}{10.0   | 7.67}                         & \multicolumn{1}{l|}{1.51   | 1.6}  & \multicolumn{1}{l|}{\cellcolor[HTML]{EFEFEF}0.99   | 0.99} & \cellcolor[HTML]{C0C0C0}25   | 15  & \multicolumn{1}{l|}{\cellcolor[HTML]{C0C0C0}11.1   | 10.1} & \multicolumn{1}{l|}{2.1   | 1.37} & \multicolumn{1}{l|}{\cellcolor[HTML]{EFEFEF}0.99   | 0.99} & \cellcolor[HTML]{C0C0C0}45   | 19   & \multicolumn{1}{l|}{18.6   | 23.4} & \multicolumn{1}{l|}{3.3   | 4.6}  & \multicolumn{1}{l|}{\cellcolor[HTML]{EFEFEF}0.99   | 0.99} & 30   | 32                           \\ \hline
\textbf{MH}                       & \multicolumn{1}{l|}{14.2   | 32.4}                         & \multicolumn{1}{l|}{2.25   | 5.03} & \multicolumn{1}{l|}{\cellcolor[HTML]{EFEFEF}0.93   | 0.93} & \cellcolor[HTML]{C0C0C0}109   | 95 & \multicolumn{1}{l|}{16.3   | 26.3}                         & \multicolumn{1}{l|}{2.37   | 4.1} & \multicolumn{1}{l|}{0.93   | 0.92}                         & \cellcolor[HTML]{C0C0C0}148   | 142 & \multicolumn{1}{l|}{47.3   | 54.1} & \multicolumn{1}{l|}{7.1   | 8.1}  & \multicolumn{1}{l|}{\cellcolor[HTML]{C0C0C0}0.94   | 0.95} & \cellcolor[HTML]{C0C0C0}143   | 110 \\ \hline
\textbf{WR}                       & \multicolumn{1}{l|}{19.7   | 28.6}                         & \multicolumn{1}{l|}{2.9   | 4.3}   & \multicolumn{1}{l|}{\cellcolor[HTML]{EFEFEF}0.94   | 0.94} & \cellcolor[HTML]{C0C0C0}133   | 94 & \multicolumn{1}{l|}{8.09   | 16.4}                         & \multicolumn{1}{l|}{1.08   | 2.7} & \multicolumn{1}{l|}{0.94   | 0.92}                         & 120   | 133                         & \multicolumn{1}{l|}{28.4   | 36.3} & \multicolumn{1}{l|}{4.4   | 5.4}  & \multicolumn{1}{l|}{\cellcolor[HTML]{C0C0C0}0.95   | 0.96} & \cellcolor[HTML]{C0C0C0}129   | 87  \\ \hline
\end{tabular}}
\caption* {\footnotesize \textnormal{AL: ALERT, BM: BeamAttack, MH: Metropolis-Hasting Modifier, WR: WIR-Random}}
\label{adv_quality_cd}
\vspace{-2.15em}
\end{table}

Table \ref{adv_quality_cs} compares the quality of adversarial examples generated from PTMs fine-tuned on human-written and LLM-generated data in the code summarization task. This table shows results similar to those of the clone detection task described above.

\begin{table}[htbp]
\centering
\caption{Comparison of the quality of adversarial examples in the code summarization task}
\vspace{-1.05em}
\resizebox{\linewidth}{!}{ % Adjusts the table to fit within the column width
\begin{tabular}{l|llll|llll|llll}
\hline
\multicolumn{1}{c|}{}                                  & \multicolumn{4}{c|}{\textbf{CodeBERT}}                                                                                                                                                                                   & \multicolumn{4}{c|}{\textbf{CodeGPT}}                                                                                                                                                                                  & \multicolumn{4}{c}{\textbf{PLBART}}                                                                                                                                                                                     \\ \cline{2-13} 
\multicolumn{1}{c|}{\multirow{-2}{*}{\textbf{Attack}}} & \multicolumn{1}{c|}{\textbf{ICR}}                          & \multicolumn{1}{c|}{\textbf{TCR}}                          & \multicolumn{1}{c|}{\textbf{ACS}}                          & \multicolumn{1}{c|}{\textbf{AED}} & \multicolumn{1}{c|}{\textbf{ICR}}                          & \multicolumn{1}{c|}{\textbf{TCR}}                        & \multicolumn{1}{c|}{\textbf{ACS}}                          & \multicolumn{1}{c|}{\textbf{AED}} & \multicolumn{1}{c|}{\textbf{ICR}}                          & \multicolumn{1}{c|}{\textbf{TCR}}                          & \multicolumn{1}{c|}{\textbf{ACS}}                          & \multicolumn{1}{c}{\textbf{AED}} \\ \hline
AL                                                      & \multicolumn{1}{l|}{16.2   | 24.5}                         & \multicolumn{1}{l|}{1.92   | 2.81}                         & \multicolumn{1}{l|}{\cellcolor[HTML]{EFEFEF}0.99   | 0.99} & \cellcolor[HTML]{C0C0C0}27   | 25 & \multicolumn{1}{l|}{\cellcolor[HTML]{EFEFEF}15.0   | 15.0} & \multicolumn{1}{l|}{1.59   | 1.6}                        & \multicolumn{1}{l|}{\cellcolor[HTML]{C0C0C0}0.99   | 0.98} & 23   | 24                         & \multicolumn{1}{l|}{2.3.4   | 27.8}                        & \multicolumn{1}{l|}{2.67   | 3.14}                         & \multicolumn{1}{l|}{\cellcolor[HTML]{EFEFEF}0.99   | 0.99} & \cellcolor[HTML]{C0C0C0}24   | 22 \\ \hline
BM                                                      & \multicolumn{1}{l|}{\cellcolor[HTML]{C0C0C0}12.3   | 5.26} & \multicolumn{1}{l|}{\cellcolor[HTML]{C0C0C0}1.3   | 0.59}  & \multicolumn{1}{l|}{\cellcolor[HTML]{EFEFEF}0.99   | 0.99} & 19   | 19                         & \multicolumn{1}{l|}{8.69   | 9.99}                         & \multicolumn{1}{l|}{\cellcolor[HTML]{EFEFEF}0.8   | 0.8} & \multicolumn{1}{l|}{0.99   | 0.99}                         & 13   | 14                         & \multicolumn{1}{l|}{16.7   | 23.1}                         & \multicolumn{1}{l|}{1.77   | 2.42}                         & \multicolumn{1}{l|}{\cellcolor[HTML]{EFEFEF}0.99   | 0.99} & 15   | 16                         \\ \hline
MH                                                      & \multicolumn{1}{l|}{\cellcolor[HTML]{EFEFEF}20.1   | 20.1} & \multicolumn{1}{l|}{1.53   | 1.76}                         & \multicolumn{1}{l|}{0.99   | 0.98}                         & 22   | 27                         & \multicolumn{1}{l|}{20.4   | 21.9}                         & \multicolumn{1}{l|}{1.44   | 1.69}                       & \multicolumn{1}{l|}{\cellcolor[HTML]{C0C0C0}0.99   | 0.98} & 23   | 25                         & \multicolumn{1}{l|}{\cellcolor[HTML]{C0C0C0}21.1   | 20.9} & \multicolumn{1}{l|}{\cellcolor[HTML]{C0C0C0}1.99   | 1.76} & \multicolumn{1}{l|}{\cellcolor[HTML]{EFEFEF}0.98   | 0.98} & \cellcolor[HTML]{C0C0C0}29   | 27 \\ \hline
WR                                                      & \multicolumn{1}{l|}{\cellcolor[HTML]{C0C0C0}9.99   | 3.62} & \multicolumn{1}{l|}{\cellcolor[HTML]{C0C0C0}1.52   | 0.58} & \multicolumn{1}{l|}{\cellcolor[HTML]{EFEFEF}0.96   | 0.96} & 65   | 66                         & \multicolumn{1}{l|}{6.35   | 7.55}                         & \multicolumn{1}{l|}{0.87   |  1.16}                      & \multicolumn{1}{l|}{\cellcolor[HTML]{C0C0C0}0.96   | 0.95} & \cellcolor[HTML]{C0C0C0}57   | 61 & \multicolumn{1}{l|}{10.8   | 12.4}                         & \multicolumn{1}{l|}{1.55   | 1.88}                         & \multicolumn{1}{l|}{\cellcolor[HTML]{EFEFEF}0.96   | 0.96} & 54   | 59                         \\ \hline
\end{tabular}}
\caption* {\footnotesize \textnormal{AL: ALERT, BM: BeamAttack, MH: Metropolis-Hasting Modifier, WR: WIR-Random}}
\label{adv_quality_cs}
\vspace{-1.95em}
\end{table}

% To quantify the quality of the generated adversarial examples, we employ several similarity metrics outlined in Section \ref{Ase PTMs}.

% \usepackage{multirow}
\begin{table*}[t]
\centering
\caption{Comparison of the quality of adversarial examples in the sentiment analysis task}
\vspace{-1.05em}
\resizebox{\linewidth}{!}{ % Adjusts the table to fit within the column width
\begin{tabular}{l|lllllll|lllllll|lllllll}
\hline
                         & \multicolumn{7}{c|}{\textbf{BERT}}                                                                                                                                                                                                                                                      & \multicolumn{7}{c|}{\textbf{RoBERTa}}                                                                                                                                                                                                                                                   & \multicolumn{7}{c}{\textbf{DistilBERT}}                                                                                                                                                                                        \\ \cline{2-22} 
\multirow{-2}{*}{\textbf{Attack}} & \multicolumn{1}{c|}{\textbf{SBERT}} & \multicolumn{1}{c|}{\textbf{LD}} & \multicolumn{1}{c|}{\textbf{LR}}                                  & \multicolumn{1}{c|}{\textbf{JSC}} & \multicolumn{1}{c|}{\textbf{DSC}} & \multicolumn{1}{c|}{\textbf{JS}} & \multicolumn{1}{c|}{\textbf{JWS}} & \multicolumn{1}{c|}{\textbf{SBERT}} & \multicolumn{1}{c|}{\textbf{LD}} & \multicolumn{1}{c|}{\textbf{LR}}                                  & \multicolumn{1}{c|}{\textbf{JSC}} & \multicolumn{1}{c|}{\textbf{DSC}} & \multicolumn{1}{c|}{\textbf{JS}} & \multicolumn{1}{c|}{\textbf{JWS}} & \multicolumn{1}{c|}{\textbf{SBERT}} & \multicolumn{1}{c|}{\textbf{LD}} & \multicolumn{1}{c|}{\textbf{LR}} & \multicolumn{1}{l|}{JSC}         & \multicolumn{1}{l|}{DSC}         & \multicolumn{1}{l|}{JS}           & JWS         \\ \hline
\textbf{BE}                       & \multicolumn{1}{l|}{0.82 | 0.76}  & \multicolumn{1}{l|}{9 | 15}      & \multicolumn{1}{l|}{0.88 | 0.81}                                  & \multicolumn{1}{l|}{0.59 | 0.46}  & \multicolumn{1}{l|}{0.71 | 0.6}   & \multicolumn{1}{l|}{0.86 | 0.81} & 0.89 | 0.85                       & \multicolumn{1}{l|}{0.89 | 0.76}  & \multicolumn{1}{l|}{6 | 15}      & \multicolumn{1}{l|}{0.91 | 0.81}                                  & \multicolumn{1}{l|}{0.68 | 0.42}  & \multicolumn{1}{l|}{0.78 | 0.57}  & \multicolumn{1}{l|}{0.88 | 0.81} & 0.91 | 0.85                       & \multicolumn{1}{l|}{0.84 | 0.77}  & \multicolumn{1}{l|}{8 | 14}      & \multicolumn{1}{l|}{0.88 | 0.82} & \multicolumn{1}{l|}{0.6 | 0.46}  & \multicolumn{1}{l|}{0.71 | 0.6}  & \multicolumn{1}{l|}{0.86 |  0.81} & 0.89 | 0.86 \\ \hline
\textbf{BA}                       & \multicolumn{1}{l|}{0.71 | 0.69}  & \multicolumn{1}{l|}{10 | 16}     & \multicolumn{1}{l|}{0.85 | 0.8}                                   & \multicolumn{1}{l|}{0.54 | 0.45}  & \multicolumn{1}{l|}{0.67 | 0.6}   & \multicolumn{1}{l|}{0.84 | 0.6}  & 0.9 | 0.81                        & \multicolumn{1}{l|}{0.75 | 0.88}  & \multicolumn{1}{l|}{9 | 15}      & \multicolumn{1}{l|}{0.87 | 0.81}                                  & \multicolumn{1}{l|}{0.58 | 0.47}  & \multicolumn{1}{l|}{0.7 | 062}    & \multicolumn{1}{l|}{0.85 | 0.82} & 0.91 |  0.89                      & \multicolumn{1}{l|}{0.74 | 0.66}  & \multicolumn{1}{l|}{9 | 16}      & \multicolumn{1}{l|}{0.86 | 0.8}  & \multicolumn{1}{l|}{0.56 | 0.45} & \multicolumn{1}{l|}{0.68 | 0.59} & \multicolumn{1}{l|}{0.85 | 0.81}  & 0.91 | 0.89 \\ \hline
\textbf{PW}                       & \multicolumn{1}{l|}{0.72 | 0.68}  & \multicolumn{1}{l|}{10 | 15}     & \multicolumn{1}{l|}{0.86 | 0.82}                                  & \multicolumn{1}{l|}{0.58 | 0.48}  & \multicolumn{1}{l|}{0.71 | 0.63}  & \multicolumn{1}{l|}{0.85 | 0.82} & 0.88 | 0.88                       & \multicolumn{1}{l|}{0.78 | 0.68}  & \multicolumn{1}{l|}{9 | 15}      & \multicolumn{1}{l|}{0.88 | 0.82}                                  & \multicolumn{1}{l|}{0.63 | 0.49}  & \multicolumn{1}{l|}{0.75 | 0.63}  & \multicolumn{1}{l|}{0.85 | 0.82} & 0.89 | 0.88                       & \multicolumn{1}{l|}{0.73 | 0.71}  & \multicolumn{1}{l|}{9 | 14}      & \multicolumn{1}{l|}{0.87 | 0.83} & \multicolumn{1}{l|}{0.59 | 0.51} & \multicolumn{1}{l|}{0.71 | 0.65} & \multicolumn{1}{l|}{0.85 | 0.83}  & {\cellcolor[HTML]{EFEFEF}0.88 | 0.88} \\ \hline
\textbf{TB}                       & \multicolumn{1}{l|}{0.83 | 0.79}  & \multicolumn{1}{l|}{3 | 4}       & \multicolumn{1}{l|}{\cellcolor[HTML]{EFEFEF}0.95 | 0.95} & \multicolumn{1}{l|}{0.67 | 0.58}  & \multicolumn{1}{l|}{0.79 | 0.72}  & \multicolumn{1}{l|}{0.93 | 0.91} & 0.95 | 0.94                       & \multicolumn{1}{l|}{0.86 | 0.79}  & \multicolumn{1}{l|}{3 | 4}       & \multicolumn{1}{l|}{\cellcolor[HTML]{EFEFEF}0.95 | 0.95} & \multicolumn{1}{l|}{0.69 | 0.57}  & \multicolumn{1}{l|}{0.8 | 0.72}   & \multicolumn{1}{l|}{0.92 | 0.9}  & 0.95 | 0.93                       & \multicolumn{1}{l|}{0.83 | 0.8}   & \multicolumn{1}{l|}{3 | 4}       & \multicolumn{1}{l|}{0.95 | 0.94} & \multicolumn{1}{l|}{0.67 | 0.59} & \multicolumn{1}{l|}{0.79 | 0.73} & \multicolumn{1}{l|}{0.92 | 0.9}   & 0.94 | 0.93 \\ \hline
\textbf{TF}                       & \multicolumn{1}{l|}{0.69 | 0.67}  & \multicolumn{1}{l|}{14 | 18}     & \multicolumn{1}{l|}{0.81 | 0.79}                                  & \multicolumn{1}{l|}{0.49 | 0.45}  & \multicolumn{1}{l|}{0.61 | 0.59}  & \multicolumn{1}{l|}{0.82 | 0.8}  & {\cellcolor[HTML]{EFEFEF}0.86 | 0.86}                       & \multicolumn{1}{l|}{0.72 | 0.68}  & \multicolumn{1}{l|}{13 | 17}     & \multicolumn{1}{l|}{0.82 | 0.8}                                   & \multicolumn{1}{l|}{0.51 | 0.47}  & \multicolumn{1}{l|}{0.63 | 0.61}  & \multicolumn{1}{l|}{0.83 | 0.81} & {\cellcolor[HTML]{A9A9A9}0.86 | 0.87}                      & \multicolumn{1}{l|}{0.69 | 0.65}  & \multicolumn{1}{l|}{13 | 18}     & \multicolumn{1}{l|}{0.82 | 0.79} & \multicolumn{1}{l|}{0.49 | 0.44} & \multicolumn{1}{l|}{0.62 | 0.58} & \multicolumn{1}{l|}{0.82 | 0.81}  & {\cellcolor[HTML]{A9A9A9}0.86 | 0.87} \\ \hline
\end{tabular}}
\caption* {\footnotesize \textnormal{BE: BERT-Attack, BA: BAEAttack, PW: PWWSAttack, TB: TextBugger, TF: TextFooler}}
\label{adv_quality_sa}
\vspace{-1.75em}
\end{table*}

% To address this question, we evaluate the quality of the adversarial examples generated for the models fine-tuned on human-written and LLM-generated data using the metrics outlined in Section \ref{Ase PTMs}.

Table \ref{adv_quality_sa} presents a detailed comparison of the quality of adversarial examples generated from PTMs fine-tuned on human-written and LLM-generated data in the sentiment analysis task. Overall, the results indicate that adversarial examples generated by PTMs fine-tuned on human-written data exhibit greater syntactic and semantic similarity to the original examples across all PTMs, with only a few exceptions. For example, the cells highlighted in light gray represent cases where the quality of the adversarial examples is similar. In contrast, the cells highlighted in dark gray represent cases where the quality of the adversarial examples is better for PTMs fine-tuned on LLM-generated data. These findings reveal that while PTMs fine-tuned on LLM-generated data demonstrate better performance prior to adversarial attacks, they are significantly more vulnerable to such attacks for sentiment analysis in code review discussions. These findings reinforce the preference for human-written data over LLM-generated data when training robust PTMs for software analytics tasks.

\begin{tcolorbox}[
    enhanced,
    attach boxed title to top left={yshift=-3mm,yshifttext=-1mm}, % Attach title to top left
    colback=mycolor_box,                 % Background color of the box
    colframe=black,                % Frame color of the box
    colbacktitle= mycolor_title,            % Background color of the title box
    coltitle=black,                % Title text color
    title=Result RQ3,            % Title text
    fonttitle=\bfseries,           % Title font
    boxed title style={size=small},% Title box style
    width=0.49\textwidth,            % Set the box width to 50% of the text width
    boxsep=1mm,                    % Reduce padding inside the box
    left=0mm,                      % Remove left padding
    right=0mm, % Remove right padding
    bottom=0mm
]

    Adversarial examples generated from PTMs fine-tuned on human-written data exhibit greater syntactic and semantic similarity to the originals than those fine-tuned on LLM-generated data, indicating a preference for human-written data in robust model training.

   % Adversarial examples generated by models fine-tuned on human-written data exhibit greater syntactic and semantic similarity to the original examples than those fine-tuned on LLM-generated data. These findings support the preference for human-written data over LLM-generated data when training or fine-tuning PTMs for software analytics tasks.

   % Overall, adversarial examples generated by models fine-tuned on human-written data exhibit greater syntactic and semantic similarity to the original examples than those fine-tuned on LLM-generated data. These findings highlight the superiority of human-written data in generating robust, high-quality adversarial examples, supporting its preference over LLM-generated data for training or fine-tuning PTMs in software analytics tasks.
    
% Pre-trained models fine-tuned on human-written data perform better than those on LLM-generated data across all the selected tasks under adversarial attacks regarding ASR and AMQ metrics, highlighting the superiority of human-written data over LLM-generated data in training or fine-tuning models for software analysis tasks.
    
    % Pre-trained models fine-tuned on LLM-generated data perform slightly better than those fine-tuned on human-written data in sentiment analysis and code summarization tasks, highlighting the strength of LLMs in natural text generation. On the other hand, in the clone detection task, models fine-tuned on human-written data significantly outperform those fine-tuned on LLM-generated data, emphasizing the limitations of LLMs in code generation.
\end{tcolorbox}
\label{PTMs_result_quality}
%\end{center}

\subsection{Study Findings}

Based on our extensive empirical study, we highlight the key takeaways below.
% This study compares the effectiveness of LLM-generated and human-written data for fine-tuning PTMs in software analytics tasks, focusing on the quality of adversarial examples and model robustness under adversarial attacks. The objective is to determine whether human-written or LLM-generated data better supports the development of robust models for these tasks. Below, we summarize the key takeaways from our experiments.

% This study examines the effectiveness of LLM-generated data for fine-tuning pre-trained models in software analytics tasks, comparing it against human-written data. The evaluation focuses on data quality in the context of adversarial attacks, assessing the robustness of PTMs that are fine-tuned on these datasets. The objective is to determine whether human-written or LLM-generated data is more suitable for developing robust models for software analytics tasks. Below, we summarize the key takeaways from our experiments. 

(1) For tasks involving natural language text, LLMs can generate data of comparable quality to human-written data for fine-tuning models. For example, PTMs fine-tuned on LLM-generated data demonstrate marginally better performance in sentiment analysis and code summarization tasks, as reflected in slight improvements in accuracy, precision, recall, BLEU-4, and F1-score metrics. However, for tasks involving only code, LLM-generated data falls short of matching the quality of human-written data in producing well-performing PTMs. For instance, in the clone detection task, PTMs fine-tuned on human-written data consistently outperform those fine-tuned on LLM-generated data across all metrics. 
    
    % These findings highlight human-written data's superiority for tasks requiring a nuanced understanding of program structure.

(2) Although PTMs fine-tuned on LLM-generated data for tasks involving natural language text performed slightly better than those fine-tuned on human-written data before adversarial attacks, their performance degraded significantly in most cases across all tasks under adversarial attacks. Notably, PTMs fine-tuned on LLM-generated data for sentiment analysis in code reviews experienced a substantial performance drop compared to those fine-tuned on human-written data despite their good pre-attack performance. For tasks involving only code, PTMs fine-tuned on human-written data consistently outperformed those fine-tuned on LLM-generated data under adversarial attacks. These findings highlight that PTMs fine-tuned on human-written data are more generalizable and robust to unseen data than those fine-tuned on LLM-generated data.
    
    % These findings underscore the advantages of using human-written data for robust model training or fine-tuning, particularly in software analytics tasks subject to adversarial attacks.

(3) Adversarial examples generated from PTMs fine-tuned on human-written data exhibit greater syntactic and semantic similarity to the original examples than those fine-tuned on LLM-generated data, as evidenced by higher similarity metric values across most models and attack techniques. Since adversarial examples can be used in adversarial fine-tuning \cite{yang2022natural, du2023extensive}, human-written data emerges as a preferable choice over LLM-generated data for improving the adversarial robustness of PTMs in software analytics.

\vspace{-1.5em}
\section{Discussion}
\label{disc}
\begin{figure}[htbp]
  \centering
  \subfigure[BERT]{\label{bert_entropy}\includegraphics[width=2.75cm, height=2.5cm]{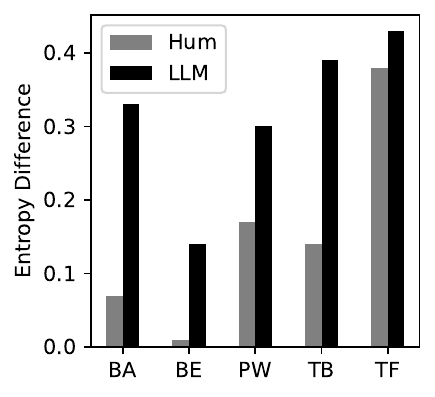}}
  \subfigure[RoBERTa]{\label{roberta_entropy}\includegraphics[width=2.75cm, height=2.5cm]{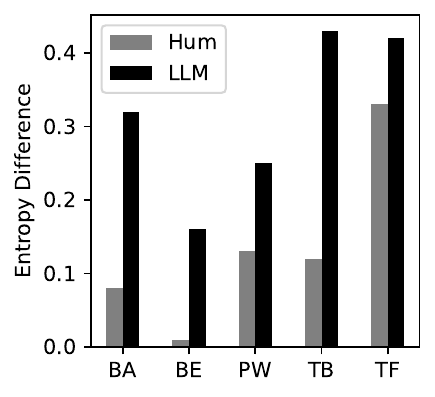}}
  \subfigure[DistilBERT]{\label{distilbert_entropy}\includegraphics[width=2.75cm, height=2.5cm]{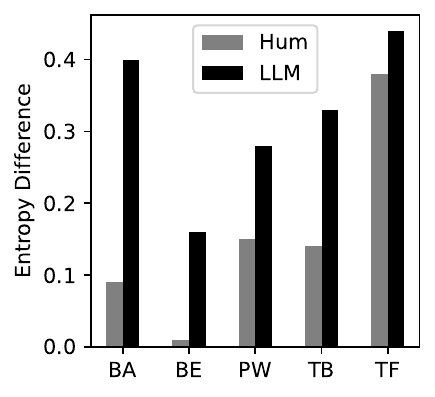}}
  \vspace{-1.05em}
  \caption{Entropy change between original and adversarial examples after adversarial attacks on PTMs fine-tuned on human-written and LLM-generated data.}
  \label{entropy}
  \Description{Comparing model performance under adversarial attacks.}
  \vspace{-1.25em}
\end{figure}

The lower robustness of PTMs fine-tuned on LLM-generated data, as observed after adversarial attacks, raises an important question: \textit{Why do these models exhibit lower robustness compared to those fine-tuned on human-written data?} To explore this, one approach is to analyze the entropy changes in human-written and LLM-generated data following adversarial attacks. In the context of sentiment analysis for code review discussions, Figure \ref{entropy} illustrates the change in entropy (e.g., word-level entropy) after adversarial attacks on PTMs. The figure clearly shows that after generating adversarial examples, entropy increases for all PTMs fine-tuned on both human-written and LLM-generated data. Notably, the increase is more pronounced for LLM-generated data than human-written data, with a substantial difference between the two. This higher entropy increase suggests that adversarial attacks introduce more uncertainty or randomness into the LLM-generated texts. As a result, PTMs fine-tuned on LLM-generated data demonstrate lower robustness than those fine-tuned on human-written data for sentiment analysis in code review discussions.

% Entropy, which measures the unpredictability or randomness in data, can provide insights into how the structure and diversity of the data influence model robustness.

Since the input for both clone detection and code summarization tasks is code, we focus on the clone detection task to investigate why PTMs fine-tuned on LLM-generated code exhibit lower robustness. ALERT, MHM, BEAMAttack, and WIR-Random attacks only perturb identifier names in the code when generating adversarial examples. As a result, we observe no change in entropy after adversarial attacks for any PTMs fine-tuned on human-written and LLM-generated code. To further compare the human-written and LLM-generated code after adversarial attacks, we calculate cyclomatic complexity to assess differences in the structural complexity of human-written and LLM-generated code. We apply the Mann-Whitney U Test to determine whether the difference in cyclomatic complexity is statistically significant. In addition, we utilize Cliff's $|\delta|$ to quantify the extent of the differences. Our statistical analysis finds that $p-\text{value}<0.05$ and Cliff's $|\delta|=0.17$, indicating the structural complexity differences among the human-written and LLM-generated codes, although the difference is not that significant due to small effect size. The differences in structural complexity may help fine-tune robust PTMs on human-written code rather than on LLM-generated code.

Since ALERT, MHM, BEAMAttack, and WIR-Random attacks only perturb identifier names, we compare the distribution of identifiers between human-written and LLM-generated codes in the clone detection task. The average number of identifiers in human-written and LLM-generated codes is 5.05 and 5.26, respectively. Additionally, the Mann-Whitney U Test finds that $p-\text{value}<0.05$ and Cliff's $|\delta|=0.1$,  suggesting the difference in the number of identifiers between human-written and LLM-generated codes. A higher number of identifiers in codes increases the likelihood of more substitutions, which may, in turn, impact the model's robustness under adversarial attacks. In conclusion, further research is needed to understand the interplay between lexical diversity and structural complexity to dictate LLMs for generating robust data in software analytics.

% In conclusion, we explore a few possible techniques to answer why PTMs fine-tuned on the LLM-generated data show lower robustness than those fine-tuned on the human-written data.

% In conclusion, we explore a few possible techniques to find the answer to why PTMs fine-tuned on the LLM-generated data show lower robustness than those fine-tuned on the human-written data. Further research is needed to better understand the interplay between lexical diversity and structural complexity to dictate LLMs for generating robust data in software analytics, as well as to explore more effective techniques in improving the capabilities of LLMs in robust data generators.

\section{Threats to Validity}
\label{threat}

% This section provides a brief overview of the construct, internal, and external threats associated with our study.

\subsection{Construct Threat}
\label{construct}
The accuracy of PTMs and the dataset preprocessing could potentially affect our empirical findings. To address this concern, we compared the robustness of PTMs, both fine-tuned on human-written and LLM-generated data under the same experimental settings as stated in \cite{yang2022natural, du2023extensive, zeng2022extensive}. The comparison between human-written and LLM-generated data relies on the assumption that both datasets are comparable in quality. Although the existing studies \cite{alam2023gptclonebench, husain2019codesearchnet} curated and validated the datasets, inherent differences in data style and structure may still affect the results. To counter this threat, we employed consistent fine-tuning strategies to address these issues across all models and tasks to ensure a fair comparison.

% The accuracy of PTMs and the dataset preprocessing could potentially affect our empirical findings. To address this concern, we compared the robustness of PTMs; both fine-tuned on human-written and LLM-generated data under the same experimental settings as stated in \cite{yang2022natural, du2023extensive, zeng2022extensive}. The comparison between human-written and LLM-generated data relies on the assumption that both datasets are comparable in quality. Although the existing studies \cite{alam2023gptclonebench, husain2019codesearchnet} curated and validated the datasets, inherent differences in data style and structure may still affect the results. Differences in fine-tuning strategies (e.g., learning rate, number of epochs, or data preprocessing) might influence the model’s robustness, which could introduce bias when comparing human-written versus LLM-generated data. We employed consistent and well-documented fine-tuning strategies to address these issues across all models and tasks to ensure a fair comparison.

\subsection{Internal Threat}
\label{internal}
% Despite our efforts to ensure robustness and validity, several internal threats may remain. 

The effectiveness of the attacks was measured using specific metrics (e.g., \%ASR, AMQ); however, they may not fully capture the semantic quality of the adversarial examples. To counter this threat, we also conducted a qualitative analysis of the adversarial examples using various similarity metrics. The effectiveness of an adversarial attack largely depends on the quality of the generated adversarial examples. If these examples are syntactically or semantically invalid, the model's failure may not accurately reflect its lack of robustness; instead, it could be a result of the model's inability to handle unrealistic inputs. To address this, we ensured that the adversarial examples maintained the semantic meaning of the original inputs. To achieve this, we utilized SOTA black-box attack methods, as described in Section \ref{Adv Attack}, which apply constraints such as minimal perturbations and semantic coherence.

% Additionally, we employed several similarity measurement metrics outlined in Section \ref{Ase PTMs} to qualitatively validate the adversarial examples. 

% The attack strategies used in this study may not encompass all possible adversarial scenarios. If more sophisticated attacks were applied, the model's robustness might be weaker than reported. We employed multiple black-box attack approaches commonly used in prior research to counter this threat and ensured they adhered to syntax and semantic-preserving constraints.

\subsection{External Threat}
\label{external}
The selected tasks- code summarization, sentiment analysis of code review comments, and clone detection—may not fully represent the entire spectrum of software analytics tasks. Other tasks, such as defect prediction or code completion, could demonstrate different levels of robustness when faced with adversarial attacks, which we did not investigate during experiments, as the LLM-generated datasets were unavailable. Despite this, we argue that the experiments using three software analytics tasks are sufficient to make our study findings generalizable. However, we encourage further research on additional tasks to compare human-written data with LLM-generated data to address the same problems, leading to a more comprehensive analysis. 

% The robustness of fine-tuned models was assessed using specific adversarial attack methods. Although these attacks are relevant and commonly used, they may not encompass all possible adversarial scenarios. Therefore, future work can explore a broader spectrum of adversarial attack methods, including more sophisticated, domain-specific strategies, to provide a comprehensive evaluation of model robustness.

%Despite the comprehensive nature of our study, specific external threats may affect the generalizability of our findings.

% More innovative or sophisticated attack methods could yield different results, potentially affecting the conclusions regarding the robustness of models fine-tuned on human-written and LLM-generated data for evaluating data quality.

\section{Related Work}
\label{RW}
Recent advancements in Large Language Models (LLMs), such as GPT-4, Deepseek-v3, Gemini, Grok 3, Claude 3.7 Sonnet, and Llama 3, have revolutionized the generation of human-like data, including natural text and code, while setting new benchmarks for a wide range of real-world applications. Despite their unprecedented successes, these models remain vulnerable to adversarial attacks in data generation, as evidenced by several studies \cite{cotroneo2023vulnerabilities, wu2023deceptprompt, wang2023adversarial}. For instance, Wu et al. \cite{wu2023deceptprompt} introduced DeceptPrompt, a method of generating adversarial natural language prompts that cause LLMs to produce code with vulnerabilities. Similarly, Cotroneo et al. \cite{cotroneo2023vulnerabilities} explored data poisoning techniques, showing that even small amounts of malicious data can compromise the security of data generated by LLMs. He et al. \cite{he2023large} proposed the SVEN technique, combining security hardening and adversarial testing to enhance the security of LLMs. Furthermore, Wang et al. \cite{wang2023adversarial} investigated adversarial attacks targeting in-context learning in LLMs, introducing the \textit{advICL} method, which manipulates input demonstrations to deceive LLMs. These studies, while contributing valuable insights into the behavior of LLMs in adversarial settings, do not directly compare the performance of LLM-generated data versus human-written data in terms of robustness to adversarial attacks in software analytics.

% Research by Allamanis et al. \cite{allamanis2024unsupervised} and Min et al. \cite{min2023beyond} emphasized the importance of evaluating the self-consistency and accuracy of models in the context of code generation.

%These findings highlight the need for robust mechanisms to ensure the security of AI-generated data.

% Several studies have focused on enhancing the robustness of LLMs in code generation. 

Several studies have focused on the quality and security of AI-generated data \cite{hamer2024just, oishwee2024large}. For example, Oishwee et al. \cite{oishwee2024large} explored the role of LLMs like ChatGPT in enhancing developer practices related to Android permissions, while Hamer et al. \cite{hamer2024just} compared the security vulnerabilities in code generated by ChatGPT and human responses from Stack Overflow. In addition, researchers assess the risks associated with data poisoning and backdoor attacks on LLMs for code across different tasks, such as code generation/completion and code summarization \cite{cotroneo2023vulnerabilities}. A large body of work has been done to evaluate the robustness of LLMs for data (e.g., natural text or code) generation \cite{yang2022natural, mastropaolo2023robustness}. Some studies investigated data privacy issues while training LLMs \cite{feng2022automated}. In the NLP domain, researchers compare the performance of humans and LLMs across various tasks \cite{li2023synthetic, bavaresco2024llms, guo2023close, nasution2024chatgpt}. Recently, Ahmed et al. \cite{ahmed2024can} explored the potential of substituting costly human subjects with more cost-effective LLM queries in evaluating code and code-related artifacts. 

While much of the existing literature focuses on LLMs' security, robustness, and efficiency in data generation and comparing their performance with human evaluators, a significant gap remains in directly comparing human-written and LLM-generated data in robust model training. This scenario raises the central question: \textit{Can LLM-generated data be effectively used for robust model training, or does human-written data provide a more reliable foundation for training robust models?} Addressing this gap, our study empirically compares the robustness of pre-trained models fine-tuned on human-written versus LLM-generated data in software analytics tasks. By evaluating adversarial examples and model performance across both datasets before and after adversarial attacks, we aim to provide insights into the effectiveness of LLMs for generating quality data, ultimately contributing a novel perspective on the comparative strength of human-written and LLM-generated data for robust model training.

\section{Conclusion}
\label{end}
This study comprehensively evaluates the quality of human-written and LLM-generated data for fine-tuning robust PTMs in software analytics. Our findings provide valuable insights into the strengths and weaknesses of both types of data regarding robustness under adversarial attacks. Before the adversarial attacks, LLM-generated data demonstrated competitive performance in natural language-based tasks such as sentiment analysis and code summarization. However, human-written data consistently outperformed LLM-generated data, particularly in tasks requiring a nuanced understanding of program structure, such as clone detection. The robustness of PTMs fine-tuned on human-written data was notably superior, as evidenced by their performance under adversarial attacks. Furthermore, adversarial examples generated from PTMs fine-tuned on human-written data were qualitatively better, showing higher syntactic and semantic coherence. These insights reaffirm the enduring value of human-written data while highlighting the challenges and opportunities in leveraging LLM-generated data to develop models that achieve high performance and robustness against adversarial attacks. Future work should focus on enhancing LLMs' capabilities to generate high-quality data that matches or surpasses human-written data, thereby supporting robust model development in software analytics.

\section*{Acknowledgement}
This research is supported in part by the Natural Sciences and Engineering Research Council of Canada (NSERC) Discovery Grants program, the Canada Foundation for Innovation's John R. Evans Leaders Fund (CFI-JELF), and by the industry-stream NSERC CREATE in Software Analytics Research (SOAR).

%This research is supported in part by the Natural Sciences and Engineering Research Council of Canada (NSERC) Discovery Grants program and by the industry-stream NSERC CREATE in Software Analytics Research (SOAR).

\bibliographystyle{ACM-Reference-Format}
\bibliography{sample-base.bib}

%%
% %% If your work has an appendix, this is the place to put it.
% \appendix

\end{document}